\documentclass[%
 reprint,
superscriptaddress,
 amsmath,amssymb,
 aps,
]{revtex4-2}

\usepackage{graphicx}% Include figure files
\usepackage{dcolumn}% Align table columns on decimal point
\usepackage{bm}% bold math
\usepackage{xcolor}
\usepackage[normalem]{ulem}

\begin{document}

%\preprint{APS/123-QED}

\title{Hybrid Symmetry Breaking for Chiral Quasi-Bound States in the Continuum}% Force line breaks with \\

\author{Yongtu Zou}
\affiliation{%
 Centre for Metamaterial Research \& Innovation, Department of Engineering, University of Exeter, \\Exeter EX4 4QF, UK
}%

\author{Zhiyao Ma}
\affiliation{
 School of Electrical and Electronic Engineering Nanyang Technological University, \\Singapore 639798, Singapore
}%

\author{Guangwei Hu}
\affiliation{
 School of Electrical and Electronic Engineering Nanyang Technological University, \\Singapore 639798, Singapore
}%

\author{Haoran Ren}
\affiliation{%
 School of Physics and Astronomy, Monash University, Melbourne, VIC 3800, Australia.
}%

\author{Stefan A. Maier}
\affiliation{%
 School of Physics and Astronomy, Monash University, Melbourne, VIC 3800, Australia.
}%
\affiliation{Department of Physics, Imperial College London, London SW7 2AZ, United Kingdom}

\author{Changxu Liu}%
 \email{C.C.Liu@exeter.ac.uk}
\affiliation{%
 Centre for Metamaterial Research \& Innovation, Department of Engineering, University of Exeter, \\Exeter EX4 4QF, UK
}%

\date{\today}% It is always \today, today,

\begin{abstract}

Chiral optical modes provide a fundamental platform for spin-selective light–matter interactions and underpin emerging applications ranging from chiral emission to polarization-controlled photonic devices.
They are typically achieved by tailoring specific structural asymmetries to directly induce circularly polarized radiation.
Here, we introduce hybrid symmetry breaking as a general route for generating and controlling chiral quasi-bound states in the continuum (qBICs).
By combining multiple orthogonal symmetry perturbations, optical chirality emerges in otherwise achiral photonic structures and evolves continuously from linear to circular polarization. 
A parity-based analysis reveals that the chiral qBICs originate from the symmetry-controlled evolution of parity-allowed radiative channels.
We demonstrate the universality of the approach across multiple BIC platforms, achieve deterministic control of the polarization state across nearly the entire Poincar\'e sphere, and further establish dynamic chirality reconfiguration in a fixed structure through a phase-change material. 
More broadly, our results demonstrate that optical chirality can emerge from symmetry engineering, providing a general framework for the design of chiral photonic states.

%\begin{description}
%You may use the \texttt{description} environment to structure your %abstract;
%\end{description}
\end{abstract}

\maketitle

\section{INTRODUCTION}

%\tableofcontents
Symmetry provides a fundamental framework for understanding physical laws and wave phenomena across diverse systems\cite{noether1983invariante,yang1996symmetry,gross1996role,kosmann2010noether}. 
In photonics, the symmetry of an optical structure governs mode formation, polarization response, and radiative coupling, thereby determining how light can be generated, confined, and manipulated.
Mirror symmetry, rotational symmetry, and inversion symmetry are central concepts in devices such as waveguides, optical cavities, photonic crystals, and metasurfaces, underpinning important effects including mode selection, polarization conversion, bound states in the continuum, and symmetry-protected transport\cite{ruter2010observation, li2015continuous, chen2015symmetry, wang2024optical,apostolico2026interaction}.  
Deliberate symmetry breaking further expands the accessible optical degrees of freedom and enables functionalities unavailable in symmetric systems, including nonlinear frequency conversion, nonreciprocal transport, topological waveguiding, and chiral light–matter interaction \cite{boyd2008nonlinear, hentschel2017chiral, chen2022multidimensional, koshelev2023nonlinear, deng2024advances,fang2012realizing, jalas2013and,  lu2014topological, ma2022topological,feng2014single, ozdemir2019parity, miri2019exceptional}.

Among these phenomena, chiral optical modes with deterministic circularly polarized radiation have recently attracted considerable attention. 
Such modes provide a compact route toward spin-selective light emission and offer opportunities for applications ranging from quantum photonics and optical communication to chiral sensing and thermal radiation engineering\cite{hentschel2017chiral,lodahl2017chiral,deng2025chiral, chen2025observation, zhang2022chiral,liu2022thermal,sun2025circularly,sun2024ultra, sinev2025chirality, both2022nanophotonic}. 
In particular, quasi-BICs provide an attractive platform for realizing chiral modes owing to their strong field confinement, high quality (Q) factors, and symmetry-engineered radiation channels\cite{hsu2016bound,koshelev2019meta,kang2023applications,wang2024optical}.

Recent studies have demonstrated chiral BICs through symmetry breaking in planar photonic structures \cite{overvig2021chiral,shi2022planar, zhang2022chiral, liu2022thermal, kuhner2023unlocking,toftul2024chiral, zhao2024spin, chen2025observation,  deng2025chiral, wang2026chiral,sun2025circularly,sun2024ultra,gorkunov2020metasurfaces}. 
Most realizations rely on breaking the out-of-plane mirror symmetry along the propagation direction, either through structural asymmetry \cite{ zhang2022chiral, gromyko2024unidirectional, sun2025circularly, deng2025chiral, chen2023observation, qin2023arbitrarily, overvig2021chiral,kang2025janus, hu2026robust, gorkunov2025substrate} or external magnetic bias\cite{mo2025brillouin,lv2024robust,zhao2024spin}. Meanwhile, it has been shown that nanostructures maintaining out-of-plane mirror symmetry can still exhibit opposite optical handedness in different momentum-space directions (wavevectors) \cite{liu2019circularly, chen2023compact, jeong2025obtuse}.
More recently, chiral modes at the $\Gamma$ point were achieved even in structures preserving vertical mirror symmetry\cite{sun2026vertical,zhang2026observation}.
In contrast to the rapidly developing understanding of symmetry-controlled radiative properties in qBICs, where a single symmetry-breaking perturbation provides a universal route to engineer and continuously tune the Q-factor via controlled leakage channels\cite{koshelev2018asymmetric}, an analogous general framework for controlling the polarization eigenstate of BICs remains absent. 
In particular, existing approaches to chiral qBICs are typically based on specific structural realizations, and do not establish a direct and universal correspondence between symmetry-breaking operations and the emergence of optical handedness. As a result, while symmetry engineering has become a powerful and predictive tool for Q-factor control, its extension to deterministic chirality control has yet to be formulated.

Here, we demonstrate a universal mechanism to endow qBICs with intrinsic optical chirality by introducing two independent symmetry-breaking operations that engineer the parity-allowed radiative channels of qBICs.
In contrast to conventional engineering, where a single symmetry perturbation governs radiative leakage and thereby controls the Q-factor, we show that an additional symmetry-breaking operation enables deterministic control of the polarization eigenstate, leading to the emergence of chiral qBICs.
To establish the generality of this principle, we implement the proposed hybrid symmetry-breaking framework in four representative and widely used BIC platforms, demonstrating that the transition from linearly polarized states to left- and right-circularly polarized modes can be universally achieved. 
The underlying physical mechanism is elucidated using a parity-based symmetry analysis, which reveals the distinct roles of the two symmetry-breaking channels in governing radiative and spin degrees of freedom.
Furthermore, by incorporating phase-change materials into the same framework, we realize dynamically reconfigurable chiral quasi-BICs, enabling continuous control of the polarization eigenstate between left- and right-circular states within a single fixed structure.

\begin{figure}[t]
    \centering
    \includegraphics[width=0.85\linewidth]{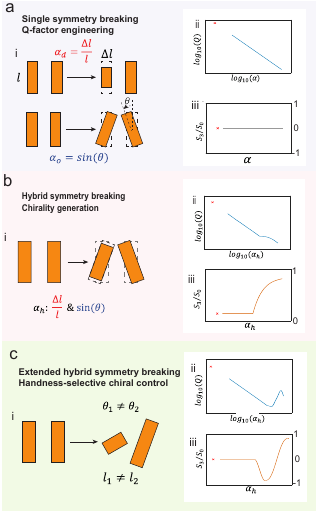}
    \caption{
    Hybrid symmetry breaking induced chiral qBICs.
    (a) Single symmetry breaking for a dimer composed of two cuboids. Dimensional asymmetry ($\alpha_d$) and orientational asymmetry ($\alpha_o$) independently generate qBICs with ($Q \propto \alpha_{d,o}^{-2}$), while the modes remain linearly polarized with ($S_3=0$).
    (b) Hybrid symmetry breaking induced chirality. A second perturbation is introduced through opposite rotations of the two cuboids after dimensional asymmetry is established. The qBIC evolves continuously from linear to circular polarization, with $S_3/S_0$ increasing from 0 to 1 as the rotational asymmetry increases.
    (c) Deterministic chirality control through extended hybrid symmetry breaking. Additional geometric degrees of freedom enable continuous tuning of the handedness of the chiral qBIC, with $S_3/S_0$ evolving from negative to positive values.
    }
    \label{Schematic}
\end{figure}

\section{RESULTS}

We first demonstrate the mechanism of chiral qBICs induced by hybrid symmetry breaking. 
As a model system, we consider a dimer structure composed of dielectric cuboids, one of the most widely used geometries. The quality factor of the resonance can be continuously tuned through symmetry breaking, following the universal scaling relation $Q \propto \alpha^{-2}$[Fig.~\ref{Schematic}a-ii], where $\alpha$ quantifies the strength of the symmetry perturbation \cite{koshelev2018asymmetric}. 
Different types of symmetry breaking can be introduced independently, including orientational symmetry breaking, where the two cuboids are rotated by opposite angles $\theta$, and dimensional symmetry breaking, where the length of one cuboid is reduced, as illustrated in Fig.~\ref{Schematic}a-i.
Under either type of symmetry breaking alone, the qBIC remains linearly polarized with the Stokes parameter $S_3=0$, indicating that the spin component is insensitive to a single perturbation channel [Fig.~\ref{Schematic}a-iii]. 

\begin{figure*}[t]
    \centering
    \includegraphics[width=0.9\linewidth]{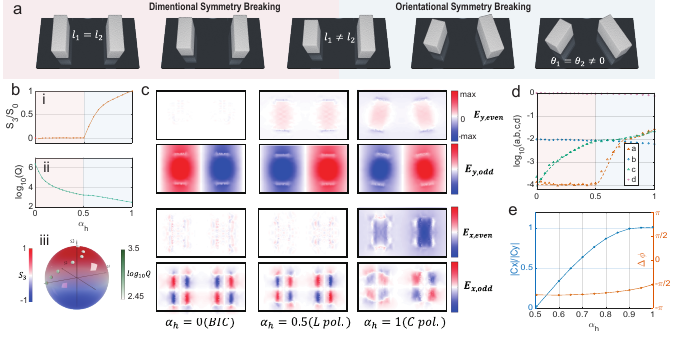}
    \caption{
    Emergence of chiral qBICs through hybrid symmetry breaking.
    (a) Geometric evolution of a rectangular dimer under dimensional (pink background) and orientational (blue background) symmetry breaking, parameterized by the hybrid asymmetry factor ($\alpha_h$).
    (b) Evolution of the (\romannumeral 1) $S_3/S_0$ and (\romannumeral 2) Q-factor. Dimensional symmetry breaking reduces Q without inducing chirality, whereas the subsequent orientational perturbation generates a chiral qBIC with continuously increasing $S_3/S_0$. (\romannumeral 3) Polarization evolution on the Poincar\'e sphere. %showing continuous tuning from linear to circular polarization.
    (c) Parity decomposition of the eigenmode fields at representative stages of the evolution with $\alpha_h=0,0.5,1$. 
    $b$ and $d$ denote the odd-parity components of $E_x$ and $E_y$, respectively, while $a$ and $c$ represent the corresponding even-parity components.
    (d) Integrated intensities of the parity-decomposed field components. Chirality emerges when the two even-parity radiative channels become comparable in magnitude. 
    (e) The amplitude ratio and phase difference of two orthogonal channels $C_x$ and $C_y$ at far field. 
    }
    \label{RectangularBar}
\end{figure*}

Remarkably, optical chiral mode emerges only when the two symmetry-breaking mechanisms are combined, as shown in Fig.~\ref{Schematic}b. 
Chirality is induced in the dimensionally perturbed qBIC only after a second perturbation is introduced through opposite rotations of the two cuboids. 
In this hybrid symmetry-breaking regime, the generated mode gradually evolves from linear polarization to a fully circularly polarized state, with $S_3/S_0$ (degree of circular polarization) continuously increasing from 0 to 1 as the rotational asymmetry is enhanced.
Furthermore, the handedness of the mode, corresponding to the sign of $S_3/S_0$, can be deterministically controlled by introducing additional geometric degrees of freedom, for example, by making the rotation angles of the two cuboids unequal [Fig.~\ref{Schematic}c].

Figure \ref{RectangularBar} quantitatively illustrates the emergence of chiral qBICs through hybrid symmetry breaking in a rectangular dielectric dimer. 
The structure consists of two dielectric cuboids with refractive index (n=3.5) suspended in a homogeneous environment, thereby excluding any symmetry breaking along the out-of-plane direction. 
The geometric dimensions are normalized to the eigenwavelength of the corresponding symmetry-protected BIC (Supplemental Material 1).

The hybrid symmetry breaking is introduced in two sequential steps [Fig.~\ref{RectangularBar}(a)].
First, dimensional asymmetry is created by shortening one cuboid, quantified by ($\alpha_d=(l_1-l_2)/l_1$), while preserving mirror symmetry with respect to the (x)-axis. 
Second, an orientational perturbation is introduced by rotating the two cuboids by opposite angles ($\theta$), characterized by ($\alpha_o=\sin\theta$). 
For convenience, the hybrid asymmetry parameter ($\alpha_h$) is defined over the entire evolution path (defined in Supplemental Material S1 and S3), with the dimensional and orientational perturbations mapped onto the intervals ($0\leq \alpha_h \leq 0.5$) and ($0.5<\alpha_h\leq1$), respectively.

The corresponding evolution of the qBIC is summarized in Fig.~\ref{RectangularBar}(b). 
During the first stage ($\alpha_h<0.5$), the quality factor decreases according to the conventional inverse-square scaling law, reflecting the gradual opening of the radiative channel. 
The normalized Stokes parameter remains zero, indicating that dimensional symmetry breaking alone is insufficient to generate optical chirality. 
Once the second perturbation is introduced ($\alpha_h>0.5$), a qualitatively different behavior emerges. 
The mode acquires a finite spin component and evolves continuously from linear to circular polarization, with $S_3/S_0$ approaching unity. 

The corresponding polarization evolution is visualized on the Poincar\'e sphere in Fig.~\ref{RectangularBar}(b)(\romannumeral 3). 
The polarization state initially remains near the equator and subsequently evolves toward the north pole as the second perturbation increases, demonstrating continuous tuning from linear to circular polarization. 
The colour purity indicates the simultaneous reduction of the quality factor during this evolution.

\begin{figure*}[t]
    \centering
    \includegraphics[width=1\linewidth]{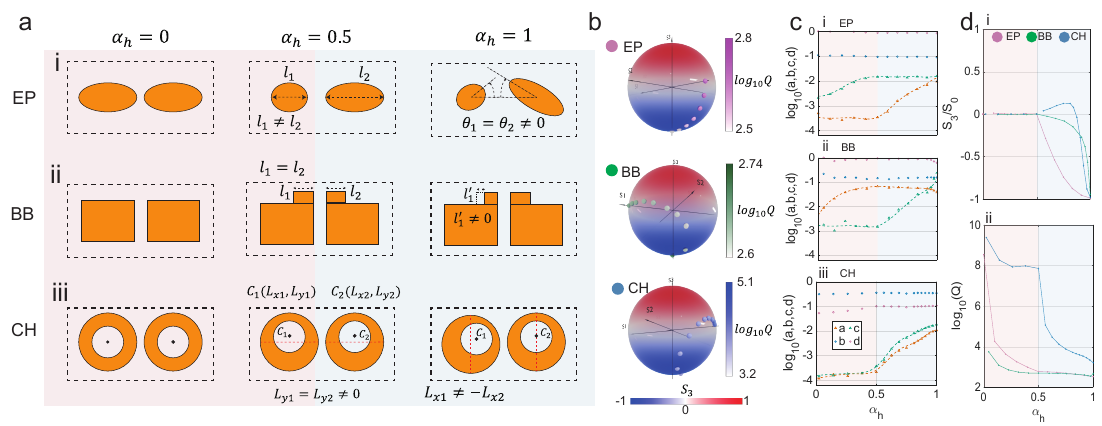}
    \caption{
    Universality of chiral qBIC generation through hybrid symmetry breaking.
    (a) Hybrid symmetry-breaking processes for three representative BIC platforms: elliptical pillars (EP), block bars (BB), and cylinders with off-centered holes (CH). 
    (b) Polarization evolution the Poincar\'e sphere  from the eigenmodes of the three structures.
    (c) Evolution of the integrated parity-decomposed field intensities $a,b,c,d$ as a function of the hybrid asymmetry factor ($\alpha_h$). 
    (d) Corresponding evolution of $S_3/S_0$ and Q-factor. 
    }
    \label{FourStructures}
\end{figure*}

To reveal the origin of the chiral qBIC, we perform a parity decomposition of the eigenmode fields. 
Because only field components with parity matching that of the radiation channel can couple to free space, the radiative properties of the qBIC are directly governed by the evolution of the even-parity field components. 
The electric fields are decomposed into even and odd components with respect to the mirror planes of  corresponding  eigenmode, 
\begin{align}
    E_{x,even}(x,y) &= \frac{E_{x}(x,y)+E_{x}(x,-y)}{2}\\
    E_{x,odd}(x,y)  &= \frac{E_{x}(x,y)-E_{x}(x,-y)}{2}\\
    E_{y,even}(x,y) &= \frac{E_{y}(x,y)+E_{y}(-x,y)}{2}\\
    E_{y,odd}(x,y)  &= \frac{E_{y}(x,y)-E_{y}(-x,y)}{2}
\end{align}
where $E_x$ ($E_y$) is decomposed with respect to the x-axis (y-axis). 
Representative field distributions are shown in Fig.~\ref{RectangularBar}(c). For the symmetry-protected BIC ($\alpha_h=0$, left column), both $E_x$ and $E_y$ possess purely odd parity, prohibiting coupling to the radiation continuum and resulting in an ideal BIC. 
Under dimensional symmetry breaking (middle column), a finite even-parity component $E_{y,even}$ emerges, opening a radiative channel and reducing the quality factor. 
However, the orthogonal component $E_{x,even}$ remains negligible, such that the emitted field remains linearly polarized. 
When the orientational perturbation is subsequently introduced, a second radiative channel associated with $E_{x,even}$ is activated. 
As its amplitude becomes comparable to that of $E_{y,even}$ (right column), the two orthogonal radiation channels interfere to generate circularly polarized emission.

This mechanism is quantified in Fig.~\ref{RectangularBar}(d), where we track the in-plane integrated intensities of the parity-decomposed fields throughout the symmetry evolution. 
Here, $b$ and $d$ denote the odd-parity components of $E_x$ and $E_y$, respectively,
while $a$ and $c$ represent the corresponding even-parity components ({Supplemental Material 1). 
Since only the even-parity components couple efficiently to the radiation continuum, the evolution of $a$ and $c$ directly determines the far-field polarization state. 
During the first perturbation stage, dimensional symmetry breaking predominantly increases $c$ while $a$ remains negligible, leading to a single dominant radiative channel and consequently linear polarization. 
Once the orientational perturbation is introduced, $a$ grows rapidly and becomes comparable to $c$, activating a second orthogonal radiative channel. 
The interference between these two channels generates a finite spin component, driving the continuous evolution from linear to circular polarization observed in Fig.~\ref{RectangularBar}(b). 
The odd-parity components $b$ and $d$ remain dominant throughout the evolution, reflecting the residual BIC character of the mode despite the increasing radiative leakage. 

To establish the connection between the parity analysis and the emitted polarization state, we further evaluate the far-field polarization amplitudes $C_x$ and $C_y$ (defined in Supplemental Material 1), as shown in Fig.~\ref{RectangularBar}(e). During the first perturbation stage, only one radiative channel is appreciably excited, rendering the relative phase between $C_x$ and $C_y$ ill-defined.
Once the second perturbation is introduced, the orthogonal polarization component emerges and the amplitude ratio $\lvert C_x/C_y \rvert$ increases continuously toward unity. 
In contrast, the phase difference $\Delta \phi$ remains close to $\pi$/2 throughout the evolution.
Consequently, the gradual balancing of the two orthogonal polarization components drives the emitted field from linear to circular.

\begin{figure}[t]
    \centering
    \includegraphics[width=0.9\linewidth]{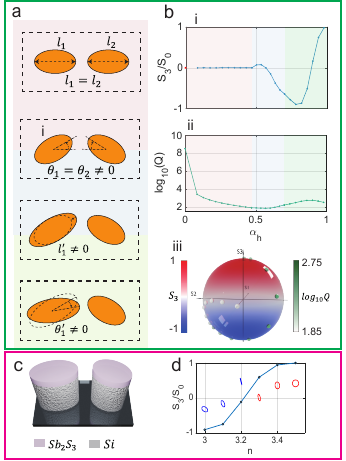}
    \caption{
    Full-Poincar\'e-sphere control and dynamic tuning of chiral qBICs.
    (a) Extended hybrid symmetry-breaking process for an elliptical-pillar dimer, consisting of orientational, dimensional, and additional single-pillar rotational perturbations.
    (b) Evolution of the normalized Stokes parameter (\romannumeral 1)$S_3/S_0$ and (\romannumeral 2)Q-factor as a function of the hybrid asymmetry factor $\alpha_h$. 
    The additional degree of freedom enables continuous tuning of $S_3/S_0$ from negative to positive values. (\romannumeral 3)Corresponding polarization evolution on the Poincar\'e sphere, demonstrating coverage of both hemispheres and near-complete access to the full spin space.
    (c) Schematic of a dynamically tunable chiral qBIC incorporating a phase-change material $Sb_2S_3$.
    (d) Polarization evolution as the refractive index of $Sb_2S_3$ is varied between its amorphous and crystalline states. The fixed geometry supports continuous tuning from right-handed circular polarization through linear polarization to left-handed circular polarization.
    }
    \label{AlldegreeOpen}
\end{figure}

To verify the generality of the mechanism, we further considered the reverse symmetry-evolution pathway, in which orientational symmetry breaking is introduced before dimensional asymmetry (Supplemental Material 2) %\textcolor{red}{and S3 and Q factor evolutions with other combinations of $\alpha_o$ and $\alpha_d$(Supplemental Material 8) } .
Similar polarization evolution and chirality generation are observed, confirming that the emergence of chiral qBICs is a generic consequence of hybrid symmetry breaking. %rather than a specific perturbation sequence.

To demonstrate the generality of the hybrid symmetry-breaking strategy, we apply it to three representative BIC platforms: 
elliptical pillars (EP), block bars (BB), and cylinders with off-centered holes (CH) [Fig.~\ref{FourStructures}(a)]. 
In each case, the symmetry evolution consists of two orthogonal perturbations that sequentially break the mirror symmetries with respect to the x and y axes. 
Details about geometry parameters and symmetry breaking processes can be found in Supplemental Material 3.
Despite the distinct geometries and perturbation pathways, all three structures exhibit a trajectory from the equator to the south pole on the Poincar\'e sphere [Fig.~\ref{FourStructures}(b)], evolving continuously from linear polarization to nearly circular polarization.

The underlying mechanism can be understood through the parity decomposition introduced above. 
Figure~\ref{FourStructures}(c) summarizes the evolution of the integrated odd-parity ($b,d$) and even-parity ($a,c$) intensities for the three structures.
For EP and BB, the first perturbation predominantly activates a single even-parity radiative channel (either $a$ or $c$), while the second perturbation drives the growth of the orthogonal channel. 
Chirality emerges when the amplitudes of the two radiative channels become comparable, enabling their interference to produce a finite spin component. 
This behavior closely follows that observed for the rectangular dimer. The CH structure exhibits a distinct evolution. During the first perturbation stage, both even-parity components $a$ and $c$ remain negligible and the mode retains an extremely high quality factor, indicating the BIC mode is still preserved.
Once the second perturbation is introduced, $a$ and $c$ increase nearly simultaneously. 
This originates from the distinct symmetry properties of the CH modes and is further discussed in Supplementary Note 4, and the far field component analysis is provided in Supplementary Note 5.}

Despite these different microscopic dynamics, the parity analysis remains fully applicable and reveals the same underlying principle: chirality is governed by the evolution of parity-allowed radiative channels in amplitude and phase space.
Accordingly, the resulting Q-factor and $S_3/S_0$ are summarized in Fig.~\ref{FourStructures}(d) as a function of $\alpha_h$.

Having established hybrid symmetry breaking as a universal route for generating chiral qBICs, we next show that extending the symmetry-breaking space enables deterministic control of the polarization state over the full Poincar\'e sphere. 
Figure~\ref{AlldegreeOpen}(a) illustrates an example based on an elliptical-pillar dimer. The evolution consists of three sequential perturbations: an initial orientational symmetry breaking ($0<\alpha_h<0.47$), followed by dimensional asymmetry ($0.47<\alpha_h<0.7$), and finally an additional rotation $\theta_1^{'}$  applied to only one pillar ($0.7<\alpha_h<1$), with more details provided in Supplemental Material 6.

The resulting polarization evolution is summarized in Fig.~\ref{AlldegreeOpen}(b). 
The first two perturbation stages reproduce the behavior discussed previously, driving the polarization state from the equator toward one pole of the Poincar\'e sphere. 
Remarkably, the third perturbation reverses the handedness of the emitted field, causing ($S_3/S_0$) to evolve continuously from negative to positive values and ultimately span nearly the entire range from -0.9 to +1 (more details in Figure S7). 
Correspondingly, the polarization trajectory traverses both hemispheres of the Poincar\'e sphere, demonstrating full control of the spin state. 
During this evolution, the quality factor exhibits a non-monotonic dependence on the geometric perturbation and reaches a maximum near $S_3=0$, highlighting the nontrivial interplay between radiative leakage and polarization chirality.

The polarization control demonstrated above relies on geometric reconfiguration. 
To enable dynamic operation in a fixed structure, we further introduce a thin layer of phase-change material $Sb_2S_3$ on top of the qBIC platform [Fig.~\ref{AlldegreeOpen}(c)]. 
Owing to the reversible phase transition between its amorphous and crystalline states, the refractive index of $Sb_2S_3$ can be continuously tuned from approximately 3.0 to 3.5\cite{dong2019wide}.
Importantly, hybrid symmetry breaking serves here as a design strategy to identify a qBIC operating point with exceptional polarization sensitivity. 
Once this geometry is established, dynamic polarization control can be achieved through a uniform refractive-index modulation, which continuously shifts the modal condition without altering the structural symmetry.
As shown in Fig.~\ref{AlldegreeOpen}(d), this refractive-index variation drives a continuous polarization evolution from right-handed circular polarization through linear polarization to left-handed circular polarization. 
In particular, a linearly polarized state is obtained near ($n\approx3.2$), where the handedness reversal occurs. The dynamic trajectory covers nearly the full Poincar\'e sphere while maintaining a fixed geometry, providing a route toward reconfigurable chiral photonic devices based on qBICs.
To the best of our knowledge, dynamic tuning of qBIC chirality of one eigenmode across both hemispheres of the Poincar\'e sphere has not been reported previously.

\section{Discussions and Conclusions}

In conclusion, we have introduced hybrid symmetry breaking as a general route for generating and controlling chiral quasi-bound states in the continuum. 
Unlike conventional approaches that rely on  geometry specific chirality, our strategy exploits the interplay between multiple symmetry perturbations to engineer the radiative channels of qBICs. 

Hybrid symmetry breaking extends the role of symmetry perturbations in BIC photonics from engineering radiative losses to continuously controlling the chirality of the emitted optical states.

The parity analysis developed here further reveals the physical origin of the chiral qBICs. 
By decomposing the eigenmodes into parity-allowed and parity-forbidden components, the emergence of optical chirality can be directly linked to the evolution of the radiative channels. 
The examples studied suggest a general design principle for chiral qBICs. 
Hybrid symmetry breaking can be engineered to selectively activate parity-allowed radiative channels associated with orthogonal field components. 

An important implication of the parity-channel framework is the trade-off between optical chirality and the quality factor. As shown in Fig. S8, Increasing $S_3/S_0$ is accompanied by a reduction in Q (more details in Supplemental Material 8) 
This behavior arises because strong chirality requires the simultaneous excitation of two parity-allowed radiative channels with comparable amplitudes, whereas a high-Q state requires weak radiative coupling. 
Consequently, enhanced chirality is generally accompanied by increased radiative leakage. This trend is also reflected in the Poincar\'e sphere evolution (Fig. 2-4), where the polarization state moves away from the equator as the quality factor decreases.

Beyond the specific structures considered here, the parity-based framework is not restricted to qBICs and may provide a general tool for understanding and engineering chiral optical modes in a broad range of photonic systems. Because the analysis is rooted in the symmetry of the eigenmodes rather than the geometric details of the structure, it offers a route for identifying and designing chiral states across diverse resonant platforms.

Furthermore, by extending the hybrid symmetry-breaking space, we demonstrate deterministic control of the polarization state across nearly the entire Poincar\'e sphere. More broadly, the results establish a route toward dynamically reconfigurable qBIC chirality spanning both hemispheres of the Poincar\'e sphere within a fixed photonic structure. 
Such capability may enable a new class of active photonic devices, including polarization-programmable nanolasers, reconfigurable chiral emitters, spin-controlled thermal sources, and dynamically tunable interfaces for chiral light–matter interactions.

\begin{acknowledgments}
G. Hu acknowledges the support from National Research Foundation of Singapore (award no. NRF-CRP31-0001), and A*STAR under its MTC YIRG Grant (Project No. M23M7c0119) and MTC IRG Grant (Project No. M24N7c0081 and No. M24N7c0087). H. Ren acknowledges the support from Australian Research Council Grant(Project No. DP220102152 and Project No. FT250100565). S. A. Maier acknowledges Lee Lucas Chair in Physics.
\end{acknowledgments}

\bibliography{ChiralBIC}% Produces the bibliography via BibTeX.

\begin{thebibliography}{55}%
\makeatletter
\providecommand \@ifxundefined [1]{%
 \@ifx{#1\undefined}
}%
\providecommand \@ifnum [1]{%
 \ifnum #1\expandafter \@firstoftwo
 \else \expandafter \@secondoftwo
 \fi
}%
\providecommand \@ifx [1]{%
 \ifx #1\expandafter \@firstoftwo
 \else \expandafter \@secondoftwo
 \fi
}%
\providecommand \natexlab [1]{#1}%
\providecommand \enquote  [1]{``#1''}%
\providecommand \bibnamefont  [1]{#1}%
\providecommand \bibfnamefont [1]{#1}%
\providecommand \citenamefont [1]{#1}%
\providecommand \href@noop [0]{\@secondoftwo}%
\providecommand \href [0]{\begingroup \@sanitize@url \@href}%
\providecommand \@href[1]{\@@startlink{#1}\@@href}%
\providecommand \@@href[1]{\endgroup#1\@@endlink}%
\providecommand \@sanitize@url [0]{\catcode `\\12\catcode `\$12\catcode `\&12\catcode `\#12\catcode `\^12\catcode `\_12\catcode `\%12\relax}%
\providecommand \@@startlink[1]{}%
\providecommand \@@endlink[0]{}%
\providecommand \url  [0]{\begingroup\@sanitize@url \@url }%
\providecommand \@url [1]{\endgroup\@href {#1}{\urlprefix }}%
\providecommand \urlprefix  [0]{URL }%
\providecommand \Eprint [0]{\href }%
\providecommand \doibase [0]{https://doi.org/}%
\providecommand \selectlanguage [0]{\@gobble}%
\providecommand \bibinfo  [0]{\@secondoftwo}%
\providecommand \bibfield  [0]{\@secondoftwo}%
\providecommand \translation [1]{[#1]}%
\providecommand \BibitemOpen [0]{}%
\providecommand \bibitemStop [0]{}%
\providecommand \bibitemNoStop [0]{.\EOS\space}%
\providecommand \EOS [0]{\spacefactor3000\relax}%
\providecommand \BibitemShut  [1]{\csname bibitem#1\endcsname}%
\let\auto@bib@innerbib\@empty
%</preamble>
\bibitem [{\citenamefont {Noether}(1983)}]{noether1983invariante}%
  \BibitemOpen
  \bibfield  {author} {\bibinfo {author} {\bibfnamefont {E.}~\bibnamefont {Noether}},\ }\bibfield  {title} {\bibinfo {title} {Invariante variationsprobleme},\ }in\ \href@noop {} {\emph {\bibinfo {booktitle} {Gesammelte abhandlungen-collected papers}}}\ (\bibinfo  {publisher} {Springer},\ \bibinfo {year} {1983})\ pp.\ \bibinfo {pages} {231--239}\BibitemShut {NoStop}%
\bibitem [{\citenamefont {Yang}(1996)}]{yang1996symmetry}%
  \BibitemOpen
  \bibfield  {author} {\bibinfo {author} {\bibfnamefont {C.~N.}\ \bibnamefont {Yang}},\ }\bibfield  {title} {\bibinfo {title} {Symmetry and physics},\ }\href@noop {} {\bibfield  {journal} {\bibinfo  {journal} {Proceedings of the American Philosophical Society}\ }\textbf {\bibinfo {volume} {140}},\ \bibinfo {pages} {267} (\bibinfo {year} {1996})}\BibitemShut {NoStop}%
\bibitem [{\citenamefont {Gross}(1996)}]{gross1996role}%
  \BibitemOpen
  \bibfield  {author} {\bibinfo {author} {\bibfnamefont {D.~J.}\ \bibnamefont {Gross}},\ }\bibfield  {title} {\bibinfo {title} {The role of symmetry in fundamental physics},\ }\href@noop {} {\bibfield  {journal} {\bibinfo  {journal} {Proceedings of the National Academy of Sciences}\ }\textbf {\bibinfo {volume} {93}},\ \bibinfo {pages} {14256} (\bibinfo {year} {1996})}\BibitemShut {NoStop}%
\bibitem [{\citenamefont {Kosmann-Schwarzbach}(2010)}]{kosmann2010noether}%
  \BibitemOpen
  \bibfield  {author} {\bibinfo {author} {\bibfnamefont {Y.}~\bibnamefont {Kosmann-Schwarzbach}},\ }\bibfield  {title} {\bibinfo {title} {The noether theorems},\ }in\ \href@noop {} {\emph {\bibinfo {booktitle} {The Noether Theorems: Invariance and Conservation Laws in the Twentieth Century}}}\ (\bibinfo  {publisher} {Springer},\ \bibinfo {year} {2010})\ pp.\ \bibinfo {pages} {55--64}\BibitemShut {NoStop}%
\bibitem [{\citenamefont {R{\"u}ter}\ \emph {et~al.}(2010)\citenamefont {R{\"u}ter}, \citenamefont {Makris}, \citenamefont {El-Ganainy}, \citenamefont {Christodoulides}, \citenamefont {Segev},\ and\ \citenamefont {Kip}}]{ruter2010observation}%
  \BibitemOpen
  \bibfield  {author} {\bibinfo {author} {\bibfnamefont {C.~E.}\ \bibnamefont {R{\"u}ter}}, \bibinfo {author} {\bibfnamefont {K.~G.}\ \bibnamefont {Makris}}, \bibinfo {author} {\bibfnamefont {R.}~\bibnamefont {El-Ganainy}}, \bibinfo {author} {\bibfnamefont {D.~N.}\ \bibnamefont {Christodoulides}}, \bibinfo {author} {\bibfnamefont {M.}~\bibnamefont {Segev}},\ and\ \bibinfo {author} {\bibfnamefont {D.}~\bibnamefont {Kip}},\ }\bibfield  {title} {\bibinfo {title} {Observation of parity--time symmetry in optics},\ }\href@noop {} {\bibfield  {journal} {\bibinfo  {journal} {Nature physics}\ }\textbf {\bibinfo {volume} {6}},\ \bibinfo {pages} {192} (\bibinfo {year} {2010})}\BibitemShut {NoStop}%
\bibitem [{\citenamefont {Li}\ \emph {et~al.}(2015)\citenamefont {Li}, \citenamefont {Chen}, \citenamefont {Pholchai}, \citenamefont {Reineke}, \citenamefont {Wong}, \citenamefont {Pun}, \citenamefont {Cheah}, \citenamefont {Zentgraf},\ and\ \citenamefont {Zhang}}]{li2015continuous}%
  \BibitemOpen
  \bibfield  {author} {\bibinfo {author} {\bibfnamefont {G.}~\bibnamefont {Li}}, \bibinfo {author} {\bibfnamefont {S.}~\bibnamefont {Chen}}, \bibinfo {author} {\bibfnamefont {N.}~\bibnamefont {Pholchai}}, \bibinfo {author} {\bibfnamefont {B.}~\bibnamefont {Reineke}}, \bibinfo {author} {\bibfnamefont {P.~W.~H.}\ \bibnamefont {Wong}}, \bibinfo {author} {\bibfnamefont {E.~Y.~B.}\ \bibnamefont {Pun}}, \bibinfo {author} {\bibfnamefont {K.~W.}\ \bibnamefont {Cheah}}, \bibinfo {author} {\bibfnamefont {T.}~\bibnamefont {Zentgraf}},\ and\ \bibinfo {author} {\bibfnamefont {S.}~\bibnamefont {Zhang}},\ }\bibfield  {title} {\bibinfo {title} {Continuous control of the nonlinearity phase for harmonic generations},\ }\href@noop {} {\bibfield  {journal} {\bibinfo  {journal} {Nature materials}\ }\textbf {\bibinfo {volume} {14}},\ \bibinfo {pages} {607} (\bibinfo {year} {2015})}\BibitemShut {NoStop}%
\bibitem [{\citenamefont {Chen}\ \emph {et~al.}(2015)\citenamefont {Chen}, \citenamefont {Zhang}, \citenamefont {Dong},\ and\ \citenamefont {Chan}}]{chen2015symmetry}%
  \BibitemOpen
  \bibfield  {author} {\bibinfo {author} {\bibfnamefont {W.-J.}\ \bibnamefont {Chen}}, \bibinfo {author} {\bibfnamefont {Z.-Q.}\ \bibnamefont {Zhang}}, \bibinfo {author} {\bibfnamefont {J.-W.}\ \bibnamefont {Dong}},\ and\ \bibinfo {author} {\bibfnamefont {C.}~\bibnamefont {Chan}},\ }\bibfield  {title} {\bibinfo {title} {Symmetry-protected transport in a pseudospin-polarized waveguide},\ }\href@noop {} {\bibfield  {journal} {\bibinfo  {journal} {Nature communications}\ }\textbf {\bibinfo {volume} {6}},\ \bibinfo {pages} {8183} (\bibinfo {year} {2015})}\BibitemShut {NoStop}%
\bibitem [{\citenamefont {Wang}\ \emph {et~al.}(2024)\citenamefont {Wang}, \citenamefont {Li}, \citenamefont {Zhao}, \citenamefont {Qian}, \citenamefont {Wang}, \citenamefont {Wang}, \citenamefont {Zhou}, \citenamefont {Han}, \citenamefont {Peng}, \citenamefont {Shi} \emph {et~al.}}]{wang2024optical}%
  \BibitemOpen
  \bibfield  {author} {\bibinfo {author} {\bibfnamefont {J.}~\bibnamefont {Wang}}, \bibinfo {author} {\bibfnamefont {P.}~\bibnamefont {Li}}, \bibinfo {author} {\bibfnamefont {X.}~\bibnamefont {Zhao}}, \bibinfo {author} {\bibfnamefont {Z.}~\bibnamefont {Qian}}, \bibinfo {author} {\bibfnamefont {X.}~\bibnamefont {Wang}}, \bibinfo {author} {\bibfnamefont {F.}~\bibnamefont {Wang}}, \bibinfo {author} {\bibfnamefont {X.}~\bibnamefont {Zhou}}, \bibinfo {author} {\bibfnamefont {D.}~\bibnamefont {Han}}, \bibinfo {author} {\bibfnamefont {C.}~\bibnamefont {Peng}}, \bibinfo {author} {\bibfnamefont {L.}~\bibnamefont {Shi}}, \emph {et~al.},\ }\bibfield  {title} {\bibinfo {title} {Optical bound states in the continuum in periodic structures: mechanisms, effects, and applications},\ }\href@noop {} {\bibfield  {journal} {\bibinfo  {journal} {Photonics Insights}\ }\textbf {\bibinfo {volume} {3}},\ \bibinfo {pages} {R01} (\bibinfo {year} {2024})}\BibitemShut {NoStop}%
\bibitem [{\citenamefont {Apostolico}\ \emph {et~al.}(2026)\citenamefont {Apostolico}, \citenamefont {Siegle}, \citenamefont {Schmid}, \citenamefont {G{\'o}mez}, \citenamefont {Schwab}, \citenamefont {Hentschel},\ and\ \citenamefont {Giessen}}]{apostolico2026interaction}%
  \BibitemOpen
  \bibfield  {author} {\bibinfo {author} {\bibfnamefont {N.}~\bibnamefont {Apostolico}}, \bibinfo {author} {\bibfnamefont {L.}~\bibnamefont {Siegle}}, \bibinfo {author} {\bibfnamefont {L.}~\bibnamefont {Schmid}}, \bibinfo {author} {\bibfnamefont {T.-D.}\ \bibnamefont {G{\'o}mez}}, \bibinfo {author} {\bibfnamefont {J.}~\bibnamefont {Schwab}}, \bibinfo {author} {\bibfnamefont {M.}~\bibnamefont {Hentschel}},\ and\ \bibinfo {author} {\bibfnamefont {H.}~\bibnamefont {Giessen}},\ }\bibfield  {title} {\bibinfo {title} {Interaction of structured light with nanostructured matter},\ }\href@noop {} {\bibfield  {journal} {\bibinfo  {journal} {Nanophotonics}\ }\textbf {\bibinfo {volume} {15}},\ \bibinfo {pages} {e70105} (\bibinfo {year} {2026})}\BibitemShut {NoStop}%
\bibitem [{\citenamefont {Boyd}\ \emph {et~al.}(2008)\citenamefont {Boyd}, \citenamefont {Gaeta},\ and\ \citenamefont {Giese}}]{boyd2008nonlinear}%
  \BibitemOpen
  \bibfield  {author} {\bibinfo {author} {\bibfnamefont {R.~W.}\ \bibnamefont {Boyd}}, \bibinfo {author} {\bibfnamefont {A.~L.}\ \bibnamefont {Gaeta}},\ and\ \bibinfo {author} {\bibfnamefont {E.}~\bibnamefont {Giese}},\ }\bibfield  {title} {\bibinfo {title} {Nonlinear optics},\ }in\ \href@noop {} {\emph {\bibinfo {booktitle} {Springer handbook of atomic, molecular, and optical physics}}}\ (\bibinfo  {publisher} {Springer},\ \bibinfo {year} {2008})\ pp.\ \bibinfo {pages} {1097--1110}\BibitemShut {NoStop}%
\bibitem [{\citenamefont {Hentschel}\ \emph {et~al.}(2017)\citenamefont {Hentschel}, \citenamefont {Sch{\"a}ferling}, \citenamefont {Duan}, \citenamefont {Giessen},\ and\ \citenamefont {Liu}}]{hentschel2017chiral}%
  \BibitemOpen
  \bibfield  {author} {\bibinfo {author} {\bibfnamefont {M.}~\bibnamefont {Hentschel}}, \bibinfo {author} {\bibfnamefont {M.}~\bibnamefont {Sch{\"a}ferling}}, \bibinfo {author} {\bibfnamefont {X.}~\bibnamefont {Duan}}, \bibinfo {author} {\bibfnamefont {H.}~\bibnamefont {Giessen}},\ and\ \bibinfo {author} {\bibfnamefont {N.}~\bibnamefont {Liu}},\ }\bibfield  {title} {\bibinfo {title} {Chiral plasmonics},\ }\href@noop {} {\bibfield  {journal} {\bibinfo  {journal} {Science advances}\ }\textbf {\bibinfo {volume} {3}},\ \bibinfo {pages} {e1602735} (\bibinfo {year} {2017})}\BibitemShut {NoStop}%
\bibitem [{\citenamefont {Chen}\ \emph {et~al.}(2022)\citenamefont {Chen}, \citenamefont {Du}, \citenamefont {Zhang}, \citenamefont {Avalos-Ovando}, \citenamefont {Wu}, \citenamefont {Xu}, \citenamefont {Liu}, \citenamefont {Okamoto}, \citenamefont {Govorov}, \citenamefont {Xiong} \emph {et~al.}}]{chen2022multidimensional}%
  \BibitemOpen
  \bibfield  {author} {\bibinfo {author} {\bibfnamefont {Y.}~\bibnamefont {Chen}}, \bibinfo {author} {\bibfnamefont {W.}~\bibnamefont {Du}}, \bibinfo {author} {\bibfnamefont {Q.}~\bibnamefont {Zhang}}, \bibinfo {author} {\bibfnamefont {O.}~\bibnamefont {Avalos-Ovando}}, \bibinfo {author} {\bibfnamefont {J.}~\bibnamefont {Wu}}, \bibinfo {author} {\bibfnamefont {Q.-H.}\ \bibnamefont {Xu}}, \bibinfo {author} {\bibfnamefont {N.}~\bibnamefont {Liu}}, \bibinfo {author} {\bibfnamefont {H.}~\bibnamefont {Okamoto}}, \bibinfo {author} {\bibfnamefont {A.~O.}\ \bibnamefont {Govorov}}, \bibinfo {author} {\bibfnamefont {Q.}~\bibnamefont {Xiong}}, \emph {et~al.},\ }\bibfield  {title} {\bibinfo {title} {Multidimensional nanoscopic chiroptics},\ }\href@noop {} {\bibfield  {journal} {\bibinfo  {journal} {Nature Reviews Physics}\ }\textbf {\bibinfo {volume} {4}},\ \bibinfo {pages} {113} (\bibinfo {year} {2022})}\BibitemShut {NoStop}%
\bibitem [{\citenamefont {Koshelev}\ \emph {et~al.}(2023)\citenamefont {Koshelev}, \citenamefont {Tonkaev},\ and\ \citenamefont {Kivshar}}]{koshelev2023nonlinear}%
  \BibitemOpen
  \bibfield  {author} {\bibinfo {author} {\bibfnamefont {K.}~\bibnamefont {Koshelev}}, \bibinfo {author} {\bibfnamefont {P.}~\bibnamefont {Tonkaev}},\ and\ \bibinfo {author} {\bibfnamefont {Y.}~\bibnamefont {Kivshar}},\ }\bibfield  {title} {\bibinfo {title} {Nonlinear chiral metaphotonics: a perspective},\ }\href@noop {} {\bibfield  {journal} {\bibinfo  {journal} {Advanced Photonics}\ }\textbf {\bibinfo {volume} {5}},\ \bibinfo {pages} {064001} (\bibinfo {year} {2023})}\BibitemShut {NoStop}%
\bibitem [{\citenamefont {Deng}\ \emph {et~al.}(2024)\citenamefont {Deng}, \citenamefont {Li}, \citenamefont {Hu}, \citenamefont {Li}, \citenamefont {Li},\ and\ \citenamefont {Deng}}]{deng2024advances}%
  \BibitemOpen
  \bibfield  {author} {\bibinfo {author} {\bibfnamefont {Q.-M.}\ \bibnamefont {Deng}}, \bibinfo {author} {\bibfnamefont {X.}~\bibnamefont {Li}}, \bibinfo {author} {\bibfnamefont {M.-X.}\ \bibnamefont {Hu}}, \bibinfo {author} {\bibfnamefont {F.-J.}\ \bibnamefont {Li}}, \bibinfo {author} {\bibfnamefont {X.}~\bibnamefont {Li}},\ and\ \bibinfo {author} {\bibfnamefont {Z.-L.}\ \bibnamefont {Deng}},\ }\bibfield  {title} {\bibinfo {title} {Advances on broadband and resonant chiral metasurfaces},\ }\href@noop {} {\bibfield  {journal} {\bibinfo  {journal} {npj Nanophotonics}\ }\textbf {\bibinfo {volume} {1}},\ \bibinfo {pages} {20} (\bibinfo {year} {2024})}\BibitemShut {NoStop}%
\bibitem [{\citenamefont {Fang}\ \emph {et~al.}(2012)\citenamefont {Fang}, \citenamefont {Yu},\ and\ \citenamefont {Fan}}]{fang2012realizing}%
  \BibitemOpen
  \bibfield  {author} {\bibinfo {author} {\bibfnamefont {K.}~\bibnamefont {Fang}}, \bibinfo {author} {\bibfnamefont {Z.}~\bibnamefont {Yu}},\ and\ \bibinfo {author} {\bibfnamefont {S.}~\bibnamefont {Fan}},\ }\bibfield  {title} {\bibinfo {title} {Realizing effective magnetic field for photons by controlling the phase of dynamic modulation},\ }\href@noop {} {\bibfield  {journal} {\bibinfo  {journal} {Nature photonics}\ }\textbf {\bibinfo {volume} {6}},\ \bibinfo {pages} {782} (\bibinfo {year} {2012})}\BibitemShut {NoStop}%
\bibitem [{\citenamefont {Jalas}\ \emph {et~al.}(2013)\citenamefont {Jalas}, \citenamefont {Petrov}, \citenamefont {Eich}, \citenamefont {Freude}, \citenamefont {Fan}, \citenamefont {Yu}, \citenamefont {Baets}, \citenamefont {Popovi{\'c}}, \citenamefont {Melloni}, \citenamefont {Joannopoulos} \emph {et~al.}}]{jalas2013and}%
  \BibitemOpen
  \bibfield  {author} {\bibinfo {author} {\bibfnamefont {D.}~\bibnamefont {Jalas}}, \bibinfo {author} {\bibfnamefont {A.}~\bibnamefont {Petrov}}, \bibinfo {author} {\bibfnamefont {M.}~\bibnamefont {Eich}}, \bibinfo {author} {\bibfnamefont {W.}~\bibnamefont {Freude}}, \bibinfo {author} {\bibfnamefont {S.}~\bibnamefont {Fan}}, \bibinfo {author} {\bibfnamefont {Z.}~\bibnamefont {Yu}}, \bibinfo {author} {\bibfnamefont {R.}~\bibnamefont {Baets}}, \bibinfo {author} {\bibfnamefont {M.}~\bibnamefont {Popovi{\'c}}}, \bibinfo {author} {\bibfnamefont {A.}~\bibnamefont {Melloni}}, \bibinfo {author} {\bibfnamefont {J.~D.}\ \bibnamefont {Joannopoulos}}, \emph {et~al.},\ }\bibfield  {title} {\bibinfo {title} {What is—and what is not—an optical isolator},\ }\href@noop {} {\bibfield  {journal} {\bibinfo  {journal} {Nature Photonics}\ }\textbf {\bibinfo {volume} {7}},\ \bibinfo {pages} {579} (\bibinfo {year} {2013})}\BibitemShut {NoStop}%
\bibitem [{\citenamefont {Lu}\ \emph {et~al.}(2014)\citenamefont {Lu}, \citenamefont {Joannopoulos},\ and\ \citenamefont {Solja{\v{c}}i{\'c}}}]{lu2014topological}%
  \BibitemOpen
  \bibfield  {author} {\bibinfo {author} {\bibfnamefont {L.}~\bibnamefont {Lu}}, \bibinfo {author} {\bibfnamefont {J.~D.}\ \bibnamefont {Joannopoulos}},\ and\ \bibinfo {author} {\bibfnamefont {M.}~\bibnamefont {Solja{\v{c}}i{\'c}}},\ }\bibfield  {title} {\bibinfo {title} {Topological photonics},\ }\href@noop {} {\bibfield  {journal} {\bibinfo  {journal} {Nature photonics}\ }\textbf {\bibinfo {volume} {8}},\ \bibinfo {pages} {821} (\bibinfo {year} {2014})}\BibitemShut {NoStop}%
\bibitem [{\citenamefont {Ma}\ \emph {et~al.}(2022)\citenamefont {Ma}, \citenamefont {Yang},\ and\ \citenamefont {Zhang}}]{ma2022topological}%
  \BibitemOpen
  \bibfield  {author} {\bibinfo {author} {\bibfnamefont {S.}~\bibnamefont {Ma}}, \bibinfo {author} {\bibfnamefont {B.}~\bibnamefont {Yang}},\ and\ \bibinfo {author} {\bibfnamefont {S.}~\bibnamefont {Zhang}},\ }\bibfield  {title} {\bibinfo {title} {Topological photonics in metamaterials},\ }\href@noop {} {\bibfield  {journal} {\bibinfo  {journal} {Photonics Insights}\ }\textbf {\bibinfo {volume} {1}},\ \bibinfo {pages} {R02} (\bibinfo {year} {2022})}\BibitemShut {NoStop}%
\bibitem [{\citenamefont {Feng}\ \emph {et~al.}(2014)\citenamefont {Feng}, \citenamefont {Wong}, \citenamefont {Ma}, \citenamefont {Wang},\ and\ \citenamefont {Zhang}}]{feng2014single}%
  \BibitemOpen
  \bibfield  {author} {\bibinfo {author} {\bibfnamefont {L.}~\bibnamefont {Feng}}, \bibinfo {author} {\bibfnamefont {Z.~J.}\ \bibnamefont {Wong}}, \bibinfo {author} {\bibfnamefont {R.-M.}\ \bibnamefont {Ma}}, \bibinfo {author} {\bibfnamefont {Y.}~\bibnamefont {Wang}},\ and\ \bibinfo {author} {\bibfnamefont {X.}~\bibnamefont {Zhang}},\ }\bibfield  {title} {\bibinfo {title} {Single-mode laser by parity-time symmetry breaking},\ }\href@noop {} {\bibfield  {journal} {\bibinfo  {journal} {Science}\ }\textbf {\bibinfo {volume} {346}},\ \bibinfo {pages} {972} (\bibinfo {year} {2014})}\BibitemShut {NoStop}%
\bibitem [{\citenamefont {{\"O}zdemir}\ \emph {et~al.}(2019)\citenamefont {{\"O}zdemir}, \citenamefont {Rotter}, \citenamefont {Nori},\ and\ \citenamefont {Yang}}]{ozdemir2019parity}%
  \BibitemOpen
  \bibfield  {author} {\bibinfo {author} {\bibfnamefont {{\c{S}}.~K.}\ \bibnamefont {{\"O}zdemir}}, \bibinfo {author} {\bibfnamefont {S.}~\bibnamefont {Rotter}}, \bibinfo {author} {\bibfnamefont {F.}~\bibnamefont {Nori}},\ and\ \bibinfo {author} {\bibfnamefont {L.}~\bibnamefont {Yang}},\ }\bibfield  {title} {\bibinfo {title} {Parity--time symmetry and exceptional points in photonics},\ }\href@noop {} {\bibfield  {journal} {\bibinfo  {journal} {Nature materials}\ }\textbf {\bibinfo {volume} {18}},\ \bibinfo {pages} {783} (\bibinfo {year} {2019})}\BibitemShut {NoStop}%
\bibitem [{\citenamefont {Miri}\ and\ \citenamefont {Alu}(2019)}]{miri2019exceptional}%
  \BibitemOpen
  \bibfield  {author} {\bibinfo {author} {\bibfnamefont {M.-A.}\ \bibnamefont {Miri}}\ and\ \bibinfo {author} {\bibfnamefont {A.}~\bibnamefont {Alu}},\ }\bibfield  {title} {\bibinfo {title} {Exceptional points in optics and photonics},\ }\href@noop {} {\bibfield  {journal} {\bibinfo  {journal} {Science}\ }\textbf {\bibinfo {volume} {363}},\ \bibinfo {pages} {eaar7709} (\bibinfo {year} {2019})}\BibitemShut {NoStop}%
\bibitem [{\citenamefont {Lodahl}\ \emph {et~al.}(2017)\citenamefont {Lodahl}, \citenamefont {Mahmoodian}, \citenamefont {Stobbe}, \citenamefont {Rauschenbeutel}, \citenamefont {Schneeweiss}, \citenamefont {Volz}, \citenamefont {Pichler},\ and\ \citenamefont {Zoller}}]{lodahl2017chiral}%
  \BibitemOpen
  \bibfield  {author} {\bibinfo {author} {\bibfnamefont {P.}~\bibnamefont {Lodahl}}, \bibinfo {author} {\bibfnamefont {S.}~\bibnamefont {Mahmoodian}}, \bibinfo {author} {\bibfnamefont {S.}~\bibnamefont {Stobbe}}, \bibinfo {author} {\bibfnamefont {A.}~\bibnamefont {Rauschenbeutel}}, \bibinfo {author} {\bibfnamefont {P.}~\bibnamefont {Schneeweiss}}, \bibinfo {author} {\bibfnamefont {J.}~\bibnamefont {Volz}}, \bibinfo {author} {\bibfnamefont {H.}~\bibnamefont {Pichler}},\ and\ \bibinfo {author} {\bibfnamefont {P.}~\bibnamefont {Zoller}},\ }\bibfield  {title} {\bibinfo {title} {Chiral quantum optics},\ }\href@noop {} {\bibfield  {journal} {\bibinfo  {journal} {Nature}\ }\textbf {\bibinfo {volume} {541}},\ \bibinfo {pages} {473} (\bibinfo {year} {2017})}\BibitemShut {NoStop}%
\bibitem [{\citenamefont {Deng}\ \emph {et~al.}(2025)\citenamefont {Deng}, \citenamefont {Jiang}, \citenamefont {Zhang}, \citenamefont {Zeng}, \citenamefont {Barkaoui}, \citenamefont {Xiao}, \citenamefont {Yu}, \citenamefont {Kivshar},\ and\ \citenamefont {Song}}]{deng2025chiral}%
  \BibitemOpen
  \bibfield  {author} {\bibinfo {author} {\bibfnamefont {H.}~\bibnamefont {Deng}}, \bibinfo {author} {\bibfnamefont {X.}~\bibnamefont {Jiang}}, \bibinfo {author} {\bibfnamefont {Y.}~\bibnamefont {Zhang}}, \bibinfo {author} {\bibfnamefont {Y.}~\bibnamefont {Zeng}}, \bibinfo {author} {\bibfnamefont {H.}~\bibnamefont {Barkaoui}}, \bibinfo {author} {\bibfnamefont {S.}~\bibnamefont {Xiao}}, \bibinfo {author} {\bibfnamefont {S.}~\bibnamefont {Yu}}, \bibinfo {author} {\bibfnamefont {Y.}~\bibnamefont {Kivshar}},\ and\ \bibinfo {author} {\bibfnamefont {Q.}~\bibnamefont {Song}},\ }\bibfield  {title} {\bibinfo {title} {Chiral lasing enabled by strong coupling},\ }\href@noop {} {\bibfield  {journal} {\bibinfo  {journal} {Science Advances}\ }\textbf {\bibinfo {volume} {11}},\ \bibinfo {pages} {eads9562} (\bibinfo {year} {2025})}\BibitemShut {NoStop}%
\bibitem [{\citenamefont {Chen}\ \emph {et~al.}(2025)\citenamefont {Chen}, \citenamefont {Wang}, \citenamefont {Si}, \citenamefont {Zhang}, \citenamefont {Yin}, \citenamefont {Chen}, \citenamefont {Lv}, \citenamefont {Tang}, \citenamefont {Zheng}, \citenamefont {Kivshar} \emph {et~al.}}]{chen2025observation}%
  \BibitemOpen
  \bibfield  {author} {\bibinfo {author} {\bibfnamefont {Y.}~\bibnamefont {Chen}}, \bibinfo {author} {\bibfnamefont {M.}~\bibnamefont {Wang}}, \bibinfo {author} {\bibfnamefont {J.}~\bibnamefont {Si}}, \bibinfo {author} {\bibfnamefont {Z.}~\bibnamefont {Zhang}}, \bibinfo {author} {\bibfnamefont {X.}~\bibnamefont {Yin}}, \bibinfo {author} {\bibfnamefont {J.}~\bibnamefont {Chen}}, \bibinfo {author} {\bibfnamefont {N.}~\bibnamefont {Lv}}, \bibinfo {author} {\bibfnamefont {C.}~\bibnamefont {Tang}}, \bibinfo {author} {\bibfnamefont {W.}~\bibnamefont {Zheng}}, \bibinfo {author} {\bibfnamefont {Y.}~\bibnamefont {Kivshar}}, \emph {et~al.},\ }\bibfield  {title} {\bibinfo {title} {Observation of chiral emission enabled by collective guided resonances},\ }\href@noop {} {\bibfield  {journal} {\bibinfo  {journal} {Nature Nanotechnology}\ }\textbf {\bibinfo {volume} {20}},\ \bibinfo {pages} {1205} (\bibinfo {year} {2025})}\BibitemShut {NoStop}%
\bibitem [{\citenamefont {Zhang}\ \emph {et~al.}(2022)\citenamefont {Zhang}, \citenamefont {Liu}, \citenamefont {Han}, \citenamefont {Kivshar},\ and\ \citenamefont {Song}}]{zhang2022chiral}%
  \BibitemOpen
  \bibfield  {author} {\bibinfo {author} {\bibfnamefont {X.}~\bibnamefont {Zhang}}, \bibinfo {author} {\bibfnamefont {Y.}~\bibnamefont {Liu}}, \bibinfo {author} {\bibfnamefont {J.}~\bibnamefont {Han}}, \bibinfo {author} {\bibfnamefont {Y.}~\bibnamefont {Kivshar}},\ and\ \bibinfo {author} {\bibfnamefont {Q.}~\bibnamefont {Song}},\ }\bibfield  {title} {\bibinfo {title} {Chiral emission from resonant metasurfaces},\ }\href@noop {} {\bibfield  {journal} {\bibinfo  {journal} {Science}\ }\textbf {\bibinfo {volume} {377}},\ \bibinfo {pages} {1215} (\bibinfo {year} {2022})}\BibitemShut {NoStop}%
\bibitem [{\citenamefont {Liu}\ \emph {et~al.}(2022)\citenamefont {Liu}, \citenamefont {Guo}, \citenamefont {Li},\ and\ \citenamefont {Fan}}]{liu2022thermal}%
  \BibitemOpen
  \bibfield  {author} {\bibinfo {author} {\bibfnamefont {T.}~\bibnamefont {Liu}}, \bibinfo {author} {\bibfnamefont {C.}~\bibnamefont {Guo}}, \bibinfo {author} {\bibfnamefont {W.}~\bibnamefont {Li}},\ and\ \bibinfo {author} {\bibfnamefont {S.}~\bibnamefont {Fan}},\ }\bibfield  {title} {\bibinfo {title} {Thermal photonics with broken symmetries},\ }\href@noop {} {\bibfield  {journal} {\bibinfo  {journal} {ELight}\ }\textbf {\bibinfo {volume} {2}},\ \bibinfo {pages} {25} (\bibinfo {year} {2022})}\BibitemShut {NoStop}%
\bibitem [{\citenamefont {Sun}\ \emph {et~al.}(2025)\citenamefont {Sun}, \citenamefont {Yang}, \citenamefont {Cai}, \citenamefont {Kivshar},\ and\ \citenamefont {Han}}]{sun2025circularly}%
  \BibitemOpen
  \bibfield  {author} {\bibinfo {author} {\bibfnamefont {K.}~\bibnamefont {Sun}}, \bibinfo {author} {\bibfnamefont {B.}~\bibnamefont {Yang}}, \bibinfo {author} {\bibfnamefont {Y.}~\bibnamefont {Cai}}, \bibinfo {author} {\bibfnamefont {Y.}~\bibnamefont {Kivshar}},\ and\ \bibinfo {author} {\bibfnamefont {Z.}~\bibnamefont {Han}},\ }\bibfield  {title} {\bibinfo {title} {Circularly polarized thermal emission driven by chiral flatbands in monoclinic metasurfaces},\ }\href@noop {} {\bibfield  {journal} {\bibinfo  {journal} {Science Advances}\ }\textbf {\bibinfo {volume} {11}},\ \bibinfo {pages} {eadw0986} (\bibinfo {year} {2025})}\BibitemShut {NoStop}%
\bibitem [{\citenamefont {Sun}\ \emph {et~al.}(2024)\citenamefont {Sun}, \citenamefont {Cai}, \citenamefont {Huang},\ and\ \citenamefont {Han}}]{sun2024ultra}%
  \BibitemOpen
  \bibfield  {author} {\bibinfo {author} {\bibfnamefont {K.}~\bibnamefont {Sun}}, \bibinfo {author} {\bibfnamefont {Y.}~\bibnamefont {Cai}}, \bibinfo {author} {\bibfnamefont {L.}~\bibnamefont {Huang}},\ and\ \bibinfo {author} {\bibfnamefont {Z.}~\bibnamefont {Han}},\ }\bibfield  {title} {\bibinfo {title} {Ultra-narrowband and rainbow-free mid-infrared thermal emitters enabled by a flat band design in distorted photonic lattices},\ }\href@noop {} {\bibfield  {journal} {\bibinfo  {journal} {Nature Communications}\ }\textbf {\bibinfo {volume} {15}},\ \bibinfo {pages} {4019} (\bibinfo {year} {2024})}\BibitemShut {NoStop}%
\bibitem [{\citenamefont {Sinev}\ \emph {et~al.}(2025)\citenamefont {Sinev}, \citenamefont {Richter}, \citenamefont {Toftul}, \citenamefont {Glebov}, \citenamefont {Koshelev}, \citenamefont {Hwang}, \citenamefont {Lancaster}, \citenamefont {Kivshar},\ and\ \citenamefont {Altug}}]{sinev2025chirality}%
  \BibitemOpen
  \bibfield  {author} {\bibinfo {author} {\bibfnamefont {I.}~\bibnamefont {Sinev}}, \bibinfo {author} {\bibfnamefont {F.~U.}\ \bibnamefont {Richter}}, \bibinfo {author} {\bibfnamefont {I.}~\bibnamefont {Toftul}}, \bibinfo {author} {\bibfnamefont {N.}~\bibnamefont {Glebov}}, \bibinfo {author} {\bibfnamefont {K.}~\bibnamefont {Koshelev}}, \bibinfo {author} {\bibfnamefont {Y.}~\bibnamefont {Hwang}}, \bibinfo {author} {\bibfnamefont {D.~G.}\ \bibnamefont {Lancaster}}, \bibinfo {author} {\bibfnamefont {Y.}~\bibnamefont {Kivshar}},\ and\ \bibinfo {author} {\bibfnamefont {H.}~\bibnamefont {Altug}},\ }\bibfield  {title} {\bibinfo {title} {Chirality encoding in resonant metasurfaces governed by lattice symmetries},\ }\href@noop {} {\bibfield  {journal} {\bibinfo  {journal} {nature communications}\ }\textbf {\bibinfo {volume} {16}},\ \bibinfo {pages} {6091} (\bibinfo {year} {2025})}\BibitemShut {NoStop}%
\bibitem [{\citenamefont {Both}\ \emph {et~al.}(2022)\citenamefont {Both}, \citenamefont {Schaferling}, \citenamefont {Sterl}, \citenamefont {Muljarov}, \citenamefont {Giessen},\ and\ \citenamefont {Weiss}}]{both2022nanophotonic}%
  \BibitemOpen
  \bibfield  {author} {\bibinfo {author} {\bibfnamefont {S.}~\bibnamefont {Both}}, \bibinfo {author} {\bibfnamefont {M.}~\bibnamefont {Schaferling}}, \bibinfo {author} {\bibfnamefont {F.}~\bibnamefont {Sterl}}, \bibinfo {author} {\bibfnamefont {E.~A.}\ \bibnamefont {Muljarov}}, \bibinfo {author} {\bibfnamefont {H.}~\bibnamefont {Giessen}},\ and\ \bibinfo {author} {\bibfnamefont {T.}~\bibnamefont {Weiss}},\ }\bibfield  {title} {\bibinfo {title} {Nanophotonic chiral sensing: how does it actually work?},\ }\href@noop {} {\bibfield  {journal} {\bibinfo  {journal} {ACS nano}\ }\textbf {\bibinfo {volume} {16}},\ \bibinfo {pages} {2822} (\bibinfo {year} {2022})}\BibitemShut {NoStop}%
\bibitem [{\citenamefont {Hsu}\ \emph {et~al.}(2016)\citenamefont {Hsu}, \citenamefont {Zhen}, \citenamefont {Stone}, \citenamefont {Joannopoulos},\ and\ \citenamefont {Solja{\v{c}}i{\'c}}}]{hsu2016bound}%
  \BibitemOpen
  \bibfield  {author} {\bibinfo {author} {\bibfnamefont {C.~W.}\ \bibnamefont {Hsu}}, \bibinfo {author} {\bibfnamefont {B.}~\bibnamefont {Zhen}}, \bibinfo {author} {\bibfnamefont {A.~D.}\ \bibnamefont {Stone}}, \bibinfo {author} {\bibfnamefont {J.~D.}\ \bibnamefont {Joannopoulos}},\ and\ \bibinfo {author} {\bibfnamefont {M.}~\bibnamefont {Solja{\v{c}}i{\'c}}},\ }\bibfield  {title} {\bibinfo {title} {Bound states in the continuum},\ }\href@noop {} {\bibfield  {journal} {\bibinfo  {journal} {Nature Reviews Materials}\ }\textbf {\bibinfo {volume} {1}},\ \bibinfo {pages} {16048} (\bibinfo {year} {2016})}\BibitemShut {NoStop}%
\bibitem [{\citenamefont {Koshelev}\ \emph {et~al.}(2019)\citenamefont {Koshelev}, \citenamefont {Bogdanov},\ and\ \citenamefont {Kivshar}}]{koshelev2019meta}%
  \BibitemOpen
  \bibfield  {author} {\bibinfo {author} {\bibfnamefont {K.}~\bibnamefont {Koshelev}}, \bibinfo {author} {\bibfnamefont {A.}~\bibnamefont {Bogdanov}},\ and\ \bibinfo {author} {\bibfnamefont {Y.}~\bibnamefont {Kivshar}},\ }\bibfield  {title} {\bibinfo {title} {Meta-optics and bound states in the continuum},\ }\href@noop {} {\bibfield  {journal} {\bibinfo  {journal} {Science Bulletin}\ }\textbf {\bibinfo {volume} {64}},\ \bibinfo {pages} {836} (\bibinfo {year} {2019})}\BibitemShut {NoStop}%
\bibitem [{\citenamefont {Kang}\ \emph {et~al.}(2023)\citenamefont {Kang}, \citenamefont {Liu}, \citenamefont {Chan},\ and\ \citenamefont {Xiao}}]{kang2023applications}%
  \BibitemOpen
  \bibfield  {author} {\bibinfo {author} {\bibfnamefont {M.}~\bibnamefont {Kang}}, \bibinfo {author} {\bibfnamefont {T.}~\bibnamefont {Liu}}, \bibinfo {author} {\bibfnamefont {C.~T.}\ \bibnamefont {Chan}},\ and\ \bibinfo {author} {\bibfnamefont {M.}~\bibnamefont {Xiao}},\ }\bibfield  {title} {\bibinfo {title} {Applications of bound states in the continuum in photonics},\ }\href@noop {} {\bibfield  {journal} {\bibinfo  {journal} {Nature Reviews Physics}\ }\textbf {\bibinfo {volume} {5}},\ \bibinfo {pages} {659} (\bibinfo {year} {2023})}\BibitemShut {NoStop}%
\bibitem [{\citenamefont {Overvig}\ \emph {et~al.}(2021)\citenamefont {Overvig}, \citenamefont {Yu},\ and\ \citenamefont {Al{\`u}}}]{overvig2021chiral}%
  \BibitemOpen
  \bibfield  {author} {\bibinfo {author} {\bibfnamefont {A.}~\bibnamefont {Overvig}}, \bibinfo {author} {\bibfnamefont {N.}~\bibnamefont {Yu}},\ and\ \bibinfo {author} {\bibfnamefont {A.}~\bibnamefont {Al{\`u}}},\ }\bibfield  {title} {\bibinfo {title} {Chiral quasi-bound states in the continuum},\ }\href@noop {} {\bibfield  {journal} {\bibinfo  {journal} {Physical Review Letters}\ }\textbf {\bibinfo {volume} {126}},\ \bibinfo {pages} {073001} (\bibinfo {year} {2021})}\BibitemShut {NoStop}%
\bibitem [{\citenamefont {Shi}\ \emph {et~al.}(2022)\citenamefont {Shi}, \citenamefont {Deng}, \citenamefont {Geng}, \citenamefont {Zeng}, \citenamefont {Zeng}, \citenamefont {Hu}, \citenamefont {Overvig}, \citenamefont {Li}, \citenamefont {Qiu}, \citenamefont {Al{\`u}} \emph {et~al.}}]{shi2022planar}%
  \BibitemOpen
  \bibfield  {author} {\bibinfo {author} {\bibfnamefont {T.}~\bibnamefont {Shi}}, \bibinfo {author} {\bibfnamefont {Z.-L.}\ \bibnamefont {Deng}}, \bibinfo {author} {\bibfnamefont {G.}~\bibnamefont {Geng}}, \bibinfo {author} {\bibfnamefont {X.}~\bibnamefont {Zeng}}, \bibinfo {author} {\bibfnamefont {Y.}~\bibnamefont {Zeng}}, \bibinfo {author} {\bibfnamefont {G.}~\bibnamefont {Hu}}, \bibinfo {author} {\bibfnamefont {A.}~\bibnamefont {Overvig}}, \bibinfo {author} {\bibfnamefont {J.}~\bibnamefont {Li}}, \bibinfo {author} {\bibfnamefont {C.-W.}\ \bibnamefont {Qiu}}, \bibinfo {author} {\bibfnamefont {A.}~\bibnamefont {Al{\`u}}}, \emph {et~al.},\ }\bibfield  {title} {\bibinfo {title} {Planar chiral metasurfaces with maximal and tunable chiroptical response driven by bound states in the continuum},\ }\href@noop {} {\bibfield  {journal} {\bibinfo  {journal} {Nature Communications}\ }\textbf {\bibinfo {volume} {13}},\ \bibinfo {pages} {4111} (\bibinfo {year} {2022})}\BibitemShut {NoStop}%
\bibitem [{\citenamefont {K{\"u}hner}\ \emph {et~al.}(2023)\citenamefont {K{\"u}hner}, \citenamefont {Wendisch}, \citenamefont {Antonov}, \citenamefont {B{\"u}rger}, \citenamefont {H{\"u}ttenhofer}, \citenamefont {de~S.~Menezes}, \citenamefont {Maier}, \citenamefont {Gorkunov}, \citenamefont {Kivshar},\ and\ \citenamefont {Tittl}}]{kuhner2023unlocking}%
  \BibitemOpen
  \bibfield  {author} {\bibinfo {author} {\bibfnamefont {L.}~\bibnamefont {K{\"u}hner}}, \bibinfo {author} {\bibfnamefont {F.~J.}\ \bibnamefont {Wendisch}}, \bibinfo {author} {\bibfnamefont {A.~A.}\ \bibnamefont {Antonov}}, \bibinfo {author} {\bibfnamefont {J.}~\bibnamefont {B{\"u}rger}}, \bibinfo {author} {\bibfnamefont {L.}~\bibnamefont {H{\"u}ttenhofer}}, \bibinfo {author} {\bibfnamefont {L.}~\bibnamefont {de~S.~Menezes}}, \bibinfo {author} {\bibfnamefont {S.~A.}\ \bibnamefont {Maier}}, \bibinfo {author} {\bibfnamefont {M.~V.}\ \bibnamefont {Gorkunov}}, \bibinfo {author} {\bibfnamefont {Y.}~\bibnamefont {Kivshar}},\ and\ \bibinfo {author} {\bibfnamefont {A.}~\bibnamefont {Tittl}},\ }\bibfield  {title} {\bibinfo {title} {Unlocking the out-of-plane dimension for photonic bound states in the continuum to achieve maximum optical chirality},\ }\href@noop {} {\bibfield  {journal} {\bibinfo  {journal} {Light: Science \& Applications}\ }\textbf {\bibinfo {volume} {12}},\ \bibinfo {pages} {250} (\bibinfo {year}
  {2023})}\BibitemShut {NoStop}%
\bibitem [{\citenamefont {Toftul}\ \emph {et~al.}(2024)\citenamefont {Toftul}, \citenamefont {Tonkaev}, \citenamefont {Koshelev}, \citenamefont {Lai}, \citenamefont {Song}, \citenamefont {Gorkunov},\ and\ \citenamefont {Kivshar}}]{toftul2024chiral}%
  \BibitemOpen
  \bibfield  {author} {\bibinfo {author} {\bibfnamefont {I.}~\bibnamefont {Toftul}}, \bibinfo {author} {\bibfnamefont {P.}~\bibnamefont {Tonkaev}}, \bibinfo {author} {\bibfnamefont {K.}~\bibnamefont {Koshelev}}, \bibinfo {author} {\bibfnamefont {F.}~\bibnamefont {Lai}}, \bibinfo {author} {\bibfnamefont {Q.}~\bibnamefont {Song}}, \bibinfo {author} {\bibfnamefont {M.}~\bibnamefont {Gorkunov}},\ and\ \bibinfo {author} {\bibfnamefont {Y.}~\bibnamefont {Kivshar}},\ }\bibfield  {title} {\bibinfo {title} {Chiral dichroism in resonant metasurfaces with monoclinic lattices},\ }\href@noop {} {\bibfield  {journal} {\bibinfo  {journal} {Physical Review Letters}\ }\textbf {\bibinfo {volume} {133}},\ \bibinfo {pages} {216901} (\bibinfo {year} {2024})}\BibitemShut {NoStop}%
\bibitem [{\citenamefont {Zhao}\ \emph {et~al.}(2024)\citenamefont {Zhao}, \citenamefont {Wang}, \citenamefont {Liu}, \citenamefont {Che}, \citenamefont {Wang}, \citenamefont {Chan}, \citenamefont {Shi},\ and\ \citenamefont {Zi}}]{zhao2024spin}%
  \BibitemOpen
  \bibfield  {author} {\bibinfo {author} {\bibfnamefont {X.}~\bibnamefont {Zhao}}, \bibinfo {author} {\bibfnamefont {J.}~\bibnamefont {Wang}}, \bibinfo {author} {\bibfnamefont {W.}~\bibnamefont {Liu}}, \bibinfo {author} {\bibfnamefont {Z.}~\bibnamefont {Che}}, \bibinfo {author} {\bibfnamefont {X.}~\bibnamefont {Wang}}, \bibinfo {author} {\bibfnamefont {C.}~\bibnamefont {Chan}}, \bibinfo {author} {\bibfnamefont {L.}~\bibnamefont {Shi}},\ and\ \bibinfo {author} {\bibfnamefont {J.}~\bibnamefont {Zi}},\ }\bibfield  {title} {\bibinfo {title} {Spin-orbit-locking chiral bound states in the continuum},\ }\href@noop {} {\bibfield  {journal} {\bibinfo  {journal} {Physical Review Letters}\ }\textbf {\bibinfo {volume} {133}},\ \bibinfo {pages} {036201} (\bibinfo {year} {2024})}\BibitemShut {NoStop}%
\bibitem [{\citenamefont {Wang}\ \emph {et~al.}(2026)\citenamefont {Wang}, \citenamefont {Lv}, \citenamefont {Zhang}, \citenamefont {Chen}, \citenamefont {Si}, \citenamefont {Chen}, \citenamefont {Tang}, \citenamefont {Yin}, \citenamefont {Liu}, \citenamefont {Xin} \emph {et~al.}}]{wang2026chiral}%
  \BibitemOpen
  \bibfield  {author} {\bibinfo {author} {\bibfnamefont {M.}~\bibnamefont {Wang}}, \bibinfo {author} {\bibfnamefont {N.}~\bibnamefont {Lv}}, \bibinfo {author} {\bibfnamefont {Z.}~\bibnamefont {Zhang}}, \bibinfo {author} {\bibfnamefont {Y.}~\bibnamefont {Chen}}, \bibinfo {author} {\bibfnamefont {J.}~\bibnamefont {Si}}, \bibinfo {author} {\bibfnamefont {J.}~\bibnamefont {Chen}}, \bibinfo {author} {\bibfnamefont {C.}~\bibnamefont {Tang}}, \bibinfo {author} {\bibfnamefont {X.}~\bibnamefont {Yin}}, \bibinfo {author} {\bibfnamefont {Z.}~\bibnamefont {Liu}}, \bibinfo {author} {\bibfnamefont {D.}~\bibnamefont {Xin}}, \emph {et~al.},\ }\bibfield  {title} {\bibinfo {title} {Chiral orbital lasing in a twisted bilayer metasurface},\ }\href@noop {} {\bibfield  {journal} {\bibinfo  {journal} {Nature Communications}\ }\textbf {\bibinfo {volume} {17}},\ \bibinfo {pages} {2369} (\bibinfo {year} {2026})}\BibitemShut {NoStop}%
\bibitem [{\citenamefont {Gorkunov}\ \emph {et~al.}(2020)\citenamefont {Gorkunov}, \citenamefont {Antonov},\ and\ \citenamefont {Kivshar}}]{gorkunov2020metasurfaces}%
  \BibitemOpen
  \bibfield  {author} {\bibinfo {author} {\bibfnamefont {M.~V.}\ \bibnamefont {Gorkunov}}, \bibinfo {author} {\bibfnamefont {A.~A.}\ \bibnamefont {Antonov}},\ and\ \bibinfo {author} {\bibfnamefont {Y.~S.}\ \bibnamefont {Kivshar}},\ }\bibfield  {title} {\bibinfo {title} {Metasurfaces with maximum chirality empowered by bound states in the continuum},\ }\href@noop {} {\bibfield  {journal} {\bibinfo  {journal} {Physical Review Letters}\ }\textbf {\bibinfo {volume} {125}},\ \bibinfo {pages} {093903} (\bibinfo {year} {2020})}\BibitemShut {NoStop}%
\bibitem [{\citenamefont {Gromyko}\ \emph {et~al.}(2024)\citenamefont {Gromyko}, \citenamefont {An}, \citenamefont {Gorelik}, \citenamefont {Xu}, \citenamefont {Lim}, \citenamefont {Lee}, \citenamefont {Tjiptoharsono}, \citenamefont {Tan}, \citenamefont {Qiu}, \citenamefont {Dong} \emph {et~al.}}]{gromyko2024unidirectional}%
  \BibitemOpen
  \bibfield  {author} {\bibinfo {author} {\bibfnamefont {D.}~\bibnamefont {Gromyko}}, \bibinfo {author} {\bibfnamefont {S.}~\bibnamefont {An}}, \bibinfo {author} {\bibfnamefont {S.}~\bibnamefont {Gorelik}}, \bibinfo {author} {\bibfnamefont {J.}~\bibnamefont {Xu}}, \bibinfo {author} {\bibfnamefont {L.~J.}\ \bibnamefont {Lim}}, \bibinfo {author} {\bibfnamefont {H.~Y.~L.}\ \bibnamefont {Lee}}, \bibinfo {author} {\bibfnamefont {F.}~\bibnamefont {Tjiptoharsono}}, \bibinfo {author} {\bibfnamefont {Z.-K.}\ \bibnamefont {Tan}}, \bibinfo {author} {\bibfnamefont {C.-W.}\ \bibnamefont {Qiu}}, \bibinfo {author} {\bibfnamefont {Z.}~\bibnamefont {Dong}}, \emph {et~al.},\ }\bibfield  {title} {\bibinfo {title} {Unidirectional chiral emission via twisted bi-layer metasurfaces},\ }\href@noop {} {\bibfield  {journal} {\bibinfo  {journal} {Nature Communications}\ }\textbf {\bibinfo {volume} {15}},\ \bibinfo {pages} {9804} (\bibinfo {year} {2024})}\BibitemShut {NoStop}%
\bibitem [{\citenamefont {Chen}\ \emph {et~al.}(2023{\natexlab{a}})\citenamefont {Chen}, \citenamefont {Deng}, \citenamefont {Sha}, \citenamefont {Chen}, \citenamefont {Wang}, \citenamefont {Chen}, \citenamefont {Wu}, \citenamefont {Chu}, \citenamefont {Kivshar}, \citenamefont {Xiao} \emph {et~al.}}]{chen2023observation}%
  \BibitemOpen
  \bibfield  {author} {\bibinfo {author} {\bibfnamefont {Y.}~\bibnamefont {Chen}}, \bibinfo {author} {\bibfnamefont {H.}~\bibnamefont {Deng}}, \bibinfo {author} {\bibfnamefont {X.}~\bibnamefont {Sha}}, \bibinfo {author} {\bibfnamefont {W.}~\bibnamefont {Chen}}, \bibinfo {author} {\bibfnamefont {R.}~\bibnamefont {Wang}}, \bibinfo {author} {\bibfnamefont {Y.-H.}\ \bibnamefont {Chen}}, \bibinfo {author} {\bibfnamefont {D.}~\bibnamefont {Wu}}, \bibinfo {author} {\bibfnamefont {J.}~\bibnamefont {Chu}}, \bibinfo {author} {\bibfnamefont {Y.~S.}\ \bibnamefont {Kivshar}}, \bibinfo {author} {\bibfnamefont {S.}~\bibnamefont {Xiao}}, \emph {et~al.},\ }\bibfield  {title} {\bibinfo {title} {Observation of intrinsic chiral bound states in the continuum},\ }\href@noop {} {\bibfield  {journal} {\bibinfo  {journal} {Nature}\ }\textbf {\bibinfo {volume} {613}},\ \bibinfo {pages} {474} (\bibinfo {year} {2023}{\natexlab{a}})}\BibitemShut {NoStop}%
\bibitem [{\citenamefont {Qin}\ \emph {et~al.}(2023)\citenamefont {Qin}, \citenamefont {Su}, \citenamefont {Liu}, \citenamefont {Zeng}, \citenamefont {Tang}, \citenamefont {Li}, \citenamefont {Shi}, \citenamefont {Huang}, \citenamefont {Qiu},\ and\ \citenamefont {Song}}]{qin2023arbitrarily}%
  \BibitemOpen
  \bibfield  {author} {\bibinfo {author} {\bibfnamefont {H.}~\bibnamefont {Qin}}, \bibinfo {author} {\bibfnamefont {Z.}~\bibnamefont {Su}}, \bibinfo {author} {\bibfnamefont {M.}~\bibnamefont {Liu}}, \bibinfo {author} {\bibfnamefont {Y.}~\bibnamefont {Zeng}}, \bibinfo {author} {\bibfnamefont {M.-C.}\ \bibnamefont {Tang}}, \bibinfo {author} {\bibfnamefont {M.}~\bibnamefont {Li}}, \bibinfo {author} {\bibfnamefont {Y.}~\bibnamefont {Shi}}, \bibinfo {author} {\bibfnamefont {W.}~\bibnamefont {Huang}}, \bibinfo {author} {\bibfnamefont {C.-W.}\ \bibnamefont {Qiu}},\ and\ \bibinfo {author} {\bibfnamefont {Q.}~\bibnamefont {Song}},\ }\bibfield  {title} {\bibinfo {title} {Arbitrarily polarized bound states in the continuum with twisted photonic crystal slabs},\ }\href@noop {} {\bibfield  {journal} {\bibinfo  {journal} {Light: Science \& Applications}\ }\textbf {\bibinfo {volume} {12}},\ \bibinfo {pages} {66} (\bibinfo {year} {2023})}\BibitemShut {NoStop}%
\bibitem [{\citenamefont {Kang}\ \emph {et~al.}(2025)\citenamefont {Kang}, \citenamefont {Xiao},\ and\ \citenamefont {Chan}}]{kang2025janus}%
  \BibitemOpen
  \bibfield  {author} {\bibinfo {author} {\bibfnamefont {M.}~\bibnamefont {Kang}}, \bibinfo {author} {\bibfnamefont {M.}~\bibnamefont {Xiao}},\ and\ \bibinfo {author} {\bibfnamefont {C.}~\bibnamefont {Chan}},\ }\bibfield  {title} {\bibinfo {title} {Janus bound states in the continuum with asymmetric topological charges},\ }\href@noop {} {\bibfield  {journal} {\bibinfo  {journal} {Physical Review Letters}\ }\textbf {\bibinfo {volume} {134}},\ \bibinfo {pages} {013805} (\bibinfo {year} {2025})}\BibitemShut {NoStop}%
\bibitem [{\citenamefont {Hu}\ \emph {et~al.}(2026)\citenamefont {Hu}, \citenamefont {Zhou}, \citenamefont {Zhang}, \citenamefont {Hu}, \citenamefont {Peng}, \citenamefont {Tang}, \citenamefont {Huang}, \citenamefont {Liu}, \citenamefont {Guan}, \citenamefont {Zhu} \emph {et~al.}}]{hu2026robust}%
  \BibitemOpen
  \bibfield  {author} {\bibinfo {author} {\bibfnamefont {H.}~\bibnamefont {Hu}}, \bibinfo {author} {\bibfnamefont {C.}~\bibnamefont {Zhou}}, \bibinfo {author} {\bibfnamefont {Y.}~\bibnamefont {Zhang}}, \bibinfo {author} {\bibfnamefont {M.-X.}\ \bibnamefont {Hu}}, \bibinfo {author} {\bibfnamefont {J.}~\bibnamefont {Peng}}, \bibinfo {author} {\bibfnamefont {H.}~\bibnamefont {Tang}}, \bibinfo {author} {\bibfnamefont {L.}~\bibnamefont {Huang}}, \bibinfo {author} {\bibfnamefont {J.}~\bibnamefont {Liu}}, \bibinfo {author} {\bibfnamefont {C.}~\bibnamefont {Guan}}, \bibinfo {author} {\bibfnamefont {Z.}~\bibnamefont {Zhu}}, \emph {et~al.},\ }\bibfield  {title} {\bibinfo {title} {Robust chirality via merging accidental bics with net zero topological charge},\ }\href@noop {} {\bibfield  {journal} {\bibinfo  {journal} {eLight}\ }\textbf {\bibinfo {volume} {6}},\ \bibinfo {pages} {12} (\bibinfo {year} {2026})}\BibitemShut {NoStop}%
\bibitem [{\citenamefont {Gorkunov}\ \emph {et~al.}(2025)\citenamefont {Gorkunov}, \citenamefont {Antonov}, \citenamefont {Mamonova}, \citenamefont {Muljarov},\ and\ \citenamefont {Kivshar}}]{gorkunov2025substrate}%
  \BibitemOpen
  \bibfield  {author} {\bibinfo {author} {\bibfnamefont {M.~V.}\ \bibnamefont {Gorkunov}}, \bibinfo {author} {\bibfnamefont {A.~A.}\ \bibnamefont {Antonov}}, \bibinfo {author} {\bibfnamefont {A.~V.}\ \bibnamefont {Mamonova}}, \bibinfo {author} {\bibfnamefont {E.~A.}\ \bibnamefont {Muljarov}},\ and\ \bibinfo {author} {\bibfnamefont {Y.}~\bibnamefont {Kivshar}},\ }\bibfield  {title} {\bibinfo {title} {Substrate-induced maximum optical chirality of planar dielectric structures},\ }\href@noop {} {\bibfield  {journal} {\bibinfo  {journal} {Advanced Optical Materials}\ }\textbf {\bibinfo {volume} {13}},\ \bibinfo {pages} {2402133} (\bibinfo {year} {2025})}\BibitemShut {NoStop}%
\bibitem [{\citenamefont {Mo}\ \emph {et~al.}(2025)\citenamefont {Mo}, \citenamefont {Zhang}, \citenamefont {Wu}, \citenamefont {Chen},\ and\ \citenamefont {Dong}}]{mo2025brillouin}%
  \BibitemOpen
  \bibfield  {author} {\bibinfo {author} {\bibfnamefont {H.-C.}\ \bibnamefont {Mo}}, \bibinfo {author} {\bibfnamefont {W.-J.}\ \bibnamefont {Zhang}}, \bibinfo {author} {\bibfnamefont {Z.-Y.}\ \bibnamefont {Wu}}, \bibinfo {author} {\bibfnamefont {X.-D.}\ \bibnamefont {Chen}},\ and\ \bibinfo {author} {\bibfnamefont {J.}~\bibnamefont {Dong}},\ }\bibfield  {title} {\bibinfo {title} {Brillouin zone folding induced spin--orbit-locking chiral bic and quasi-bic},\ }\href@noop {} {\bibfield  {journal} {\bibinfo  {journal} {ACS Photonics}\ }\textbf {\bibinfo {volume} {12}},\ \bibinfo {pages} {2795} (\bibinfo {year} {2025})}\BibitemShut {NoStop}%
\bibitem [{\citenamefont {Lv}\ \emph {et~al.}(2024)\citenamefont {Lv}, \citenamefont {Qin}, \citenamefont {Su}, \citenamefont {Zhang}, \citenamefont {Huang}, \citenamefont {Shi}, \citenamefont {Li}, \citenamefont {Genevet},\ and\ \citenamefont {Song}}]{lv2024robust}%
  \BibitemOpen
  \bibfield  {author} {\bibinfo {author} {\bibfnamefont {W.}~\bibnamefont {Lv}}, \bibinfo {author} {\bibfnamefont {H.}~\bibnamefont {Qin}}, \bibinfo {author} {\bibfnamefont {Z.}~\bibnamefont {Su}}, \bibinfo {author} {\bibfnamefont {C.}~\bibnamefont {Zhang}}, \bibinfo {author} {\bibfnamefont {J.}~\bibnamefont {Huang}}, \bibinfo {author} {\bibfnamefont {Y.}~\bibnamefont {Shi}}, \bibinfo {author} {\bibfnamefont {B.}~\bibnamefont {Li}}, \bibinfo {author} {\bibfnamefont {P.}~\bibnamefont {Genevet}},\ and\ \bibinfo {author} {\bibfnamefont {Q.}~\bibnamefont {Song}},\ }\bibfield  {title} {\bibinfo {title} {Robust generation of intrinsic c points with magneto-optical bound states in the continuum},\ }\href@noop {} {\bibfield  {journal} {\bibinfo  {journal} {Science advances}\ }\textbf {\bibinfo {volume} {10}},\ \bibinfo {pages} {eads0157} (\bibinfo {year} {2024})}\BibitemShut {NoStop}%
\bibitem [{\citenamefont {Liu}\ \emph {et~al.}(2019)\citenamefont {Liu}, \citenamefont {Wang}, \citenamefont {Zhang}, \citenamefont {Wang}, \citenamefont {Zhao}, \citenamefont {Guan}, \citenamefont {Liu}, \citenamefont {Shi},\ and\ \citenamefont {Zi}}]{liu2019circularly}%
  \BibitemOpen
  \bibfield  {author} {\bibinfo {author} {\bibfnamefont {W.}~\bibnamefont {Liu}}, \bibinfo {author} {\bibfnamefont {B.}~\bibnamefont {Wang}}, \bibinfo {author} {\bibfnamefont {Y.}~\bibnamefont {Zhang}}, \bibinfo {author} {\bibfnamefont {J.}~\bibnamefont {Wang}}, \bibinfo {author} {\bibfnamefont {M.}~\bibnamefont {Zhao}}, \bibinfo {author} {\bibfnamefont {F.}~\bibnamefont {Guan}}, \bibinfo {author} {\bibfnamefont {X.}~\bibnamefont {Liu}}, \bibinfo {author} {\bibfnamefont {L.}~\bibnamefont {Shi}},\ and\ \bibinfo {author} {\bibfnamefont {J.}~\bibnamefont {Zi}},\ }\bibfield  {title} {\bibinfo {title} {Circularly polarized states spawning from bound states in the continuum},\ }\href@noop {} {\bibfield  {journal} {\bibinfo  {journal} {Physical Review Letters}\ }\textbf {\bibinfo {volume} {123}},\ \bibinfo {pages} {116104} (\bibinfo {year} {2019})}\BibitemShut {NoStop}%
\bibitem [{\citenamefont {Chen}\ \emph {et~al.}(2023{\natexlab{b}})\citenamefont {Chen}, \citenamefont {Feng}, \citenamefont {Huang}, \citenamefont {Chen}, \citenamefont {Su}, \citenamefont {Ghosh}, \citenamefont {Hou}, \citenamefont {Xiong},\ and\ \citenamefont {Qiu}}]{chen2023compact}%
  \BibitemOpen
  \bibfield  {author} {\bibinfo {author} {\bibfnamefont {Y.}~\bibnamefont {Chen}}, \bibinfo {author} {\bibfnamefont {J.}~\bibnamefont {Feng}}, \bibinfo {author} {\bibfnamefont {Y.}~\bibnamefont {Huang}}, \bibinfo {author} {\bibfnamefont {W.}~\bibnamefont {Chen}}, \bibinfo {author} {\bibfnamefont {R.}~\bibnamefont {Su}}, \bibinfo {author} {\bibfnamefont {S.}~\bibnamefont {Ghosh}}, \bibinfo {author} {\bibfnamefont {Y.}~\bibnamefont {Hou}}, \bibinfo {author} {\bibfnamefont {Q.}~\bibnamefont {Xiong}},\ and\ \bibinfo {author} {\bibfnamefont {C.-W.}\ \bibnamefont {Qiu}},\ }\bibfield  {title} {\bibinfo {title} {Compact spin-valley-locked perovskite emission},\ }\href@noop {} {\bibfield  {journal} {\bibinfo  {journal} {Nature Materials}\ }\textbf {\bibinfo {volume} {22}},\ \bibinfo {pages} {1065} (\bibinfo {year} {2023}{\natexlab{b}})}\BibitemShut {NoStop}%
\bibitem [{\citenamefont {Jeong}\ \emph {et~al.}(2025)\citenamefont {Jeong}, \citenamefont {Lee}, \citenamefont {Kim}, \citenamefont {Gong}, \citenamefont {Fang}, \citenamefont {Yang}, \citenamefont {Chae},\ and\ \citenamefont {Rho}}]{jeong2025obtuse}%
  \BibitemOpen
  \bibfield  {author} {\bibinfo {author} {\bibfnamefont {M.}~\bibnamefont {Jeong}}, \bibinfo {author} {\bibfnamefont {J.}~\bibnamefont {Lee}}, \bibinfo {author} {\bibfnamefont {S.}~\bibnamefont {Kim}}, \bibinfo {author} {\bibfnamefont {X.}~\bibnamefont {Gong}}, \bibinfo {author} {\bibfnamefont {R.}~\bibnamefont {Fang}}, \bibinfo {author} {\bibfnamefont {Y.}~\bibnamefont {Yang}}, \bibinfo {author} {\bibfnamefont {S.~H.}\ \bibnamefont {Chae}},\ and\ \bibinfo {author} {\bibfnamefont {J.}~\bibnamefont {Rho}},\ }\bibfield  {title} {\bibinfo {title} {Obtuse-angled separation of chiral resonances with planar asymmetry--induced tunability of quality factors},\ }\href@noop {} {\bibfield  {journal} {\bibinfo  {journal} {Science advances}\ }\textbf {\bibinfo {volume} {11}},\ \bibinfo {pages} {eadu4875} (\bibinfo {year} {2025})}\BibitemShut {NoStop}%
\bibitem [{\citenamefont {Sun}\ \emph {et~al.}(2026)\citenamefont {Sun}, \citenamefont {Hu}, \citenamefont {Liu}, \citenamefont {Chen}, \citenamefont {Gromyko}, \citenamefont {Shi}, \citenamefont {Wu}, \citenamefont {Jin}, \citenamefont {He},\ and\ \citenamefont {Qiu}}]{sun2026vertical}%
  \BibitemOpen
  \bibfield  {author} {\bibinfo {author} {\bibfnamefont {Y.}~\bibnamefont {Sun}}, \bibinfo {author} {\bibfnamefont {Z.}~\bibnamefont {Hu}}, \bibinfo {author} {\bibfnamefont {M.}~\bibnamefont {Liu}}, \bibinfo {author} {\bibfnamefont {J.}~\bibnamefont {Chen}}, \bibinfo {author} {\bibfnamefont {D.}~\bibnamefont {Gromyko}}, \bibinfo {author} {\bibfnamefont {K.}~\bibnamefont {Shi}}, \bibinfo {author} {\bibfnamefont {L.}~\bibnamefont {Wu}}, \bibinfo {author} {\bibfnamefont {Y.}~\bibnamefont {Jin}}, \bibinfo {author} {\bibfnamefont {S.}~\bibnamefont {He}},\ and\ \bibinfo {author} {\bibfnamefont {C.-W.}\ \bibnamefont {Qiu}},\ }\bibfield  {title} {\bibinfo {title} {Vertical chiral emission from an intrinsically achiral metasurface enabled with anisotropic continuum},\ }\href@noop {} {\bibfield  {journal} {\bibinfo  {journal} {Nature Communications}\ } (\bibinfo {year} {2026})}\BibitemShut {NoStop}%
\bibitem [{\citenamefont {Zhang}\ \emph {et~al.}(2026)\citenamefont {Zhang}, \citenamefont {Wang}, \citenamefont {Wu}, \citenamefont {Liu}, \citenamefont {Zhao}, \citenamefont {Li},\ and\ \citenamefont {Liu}}]{zhang2026observation}%
  \BibitemOpen
  \bibfield  {author} {\bibinfo {author} {\bibfnamefont {R.}~\bibnamefont {Zhang}}, \bibinfo {author} {\bibfnamefont {L.}~\bibnamefont {Wang}}, \bibinfo {author} {\bibfnamefont {Z.-M.}\ \bibnamefont {Wu}}, \bibinfo {author} {\bibfnamefont {Y.-X.}\ \bibnamefont {Liu}}, \bibinfo {author} {\bibfnamefont {H.}~\bibnamefont {Zhao}}, \bibinfo {author} {\bibfnamefont {X.-C.}\ \bibnamefont {Li}},\ and\ \bibinfo {author} {\bibfnamefont {Q.~H.}\ \bibnamefont {Liu}},\ }\bibfield  {title} {\bibinfo {title} {Observation of terahertz planar chirality enabled by c-points in accidental bound states in the continuum},\ }\href@noop {} {\bibfield  {journal} {\bibinfo  {journal} {Laser \& Photonics Reviews}\ }\textbf {\bibinfo {volume} {20}},\ \bibinfo {pages} {e01023} (\bibinfo {year} {2026})}\BibitemShut {NoStop}%
\bibitem [{\citenamefont {Koshelev}\ \emph {et~al.}(2018)\citenamefont {Koshelev}, \citenamefont {Lepeshov}, \citenamefont {Liu}, \citenamefont {Bogdanov},\ and\ \citenamefont {Kivshar}}]{koshelev2018asymmetric}%
  \BibitemOpen
  \bibfield  {author} {\bibinfo {author} {\bibfnamefont {K.}~\bibnamefont {Koshelev}}, \bibinfo {author} {\bibfnamefont {S.}~\bibnamefont {Lepeshov}}, \bibinfo {author} {\bibfnamefont {M.}~\bibnamefont {Liu}}, \bibinfo {author} {\bibfnamefont {A.}~\bibnamefont {Bogdanov}},\ and\ \bibinfo {author} {\bibfnamefont {Y.}~\bibnamefont {Kivshar}},\ }\bibfield  {title} {\bibinfo {title} {Asymmetric metasurfaces with high-q resonances governed by bound states in the continuum},\ }\href@noop {} {\bibfield  {journal} {\bibinfo  {journal} {Physical review letters}\ }\textbf {\bibinfo {volume} {121}},\ \bibinfo {pages} {193903} (\bibinfo {year} {2018})}\BibitemShut {NoStop}%
\bibitem [{\citenamefont {Dong}\ \emph {et~al.}(2019)\citenamefont {Dong}, \citenamefont {Liu}, \citenamefont {Behera}, \citenamefont {Lu}, \citenamefont {Ng}, \citenamefont {Sreekanth}, \citenamefont {Zhou}, \citenamefont {Yang},\ and\ \citenamefont {Simpson}}]{dong2019wide}%
  \BibitemOpen
  \bibfield  {author} {\bibinfo {author} {\bibfnamefont {W.}~\bibnamefont {Dong}}, \bibinfo {author} {\bibfnamefont {H.}~\bibnamefont {Liu}}, \bibinfo {author} {\bibfnamefont {J.~K.}\ \bibnamefont {Behera}}, \bibinfo {author} {\bibfnamefont {L.}~\bibnamefont {Lu}}, \bibinfo {author} {\bibfnamefont {R.~J.}\ \bibnamefont {Ng}}, \bibinfo {author} {\bibfnamefont {K.~V.}\ \bibnamefont {Sreekanth}}, \bibinfo {author} {\bibfnamefont {X.}~\bibnamefont {Zhou}}, \bibinfo {author} {\bibfnamefont {J.~K.}\ \bibnamefont {Yang}},\ and\ \bibinfo {author} {\bibfnamefont {R.~E.}\ \bibnamefont {Simpson}},\ }\bibfield  {title} {\bibinfo {title} {Wide bandgap phase change material tuned visible photonics},\ }\href@noop {} {\bibfield  {journal} {\bibinfo  {journal} {Advanced Functional Materials}\ }\textbf {\bibinfo {volume} {29}},\ \bibinfo {pages} {1806181} (\bibinfo {year} {2019})}\BibitemShut {NoStop}%
\end{thebibliography}%

%\appendix 
%\input{Suppl}

\end{document}

\fontsize{11.5pt}{14pt}\selectfont
\renewcommand{\refname}{Supplemental References}
%\maketitle

\newpage
\subsection*{Note S1: The illustration of simulation method and geometrical configuration of rectangular bar dimer}

\begin{figure}[H]
    \centering
    \includegraphics[width=0.8\linewidth]{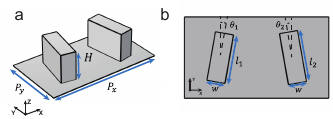}
    \caption{\textbf{Geometric configuration of cuboid dimer.} (a) oblique view (b) top view. The geometric parameters are labeled in the figure.}
    \label{fig:simu}
\end{figure}

The geometry configuration of the rectangular bar dimer is shown in Figure \ref{fig:simu}. The material of the two rectangular bars is set as silicon with a refractive index of 3.5, and the surrounding environment is air. Here, the eigenmode of the dimer is solved by COMSOL, with boundary conditions in the X/Y directions being periodic and perfectly matched in the Z direction.

The resonance wavelength of the final chiral qBIC mode($\alpha=1$) is $\lambda_0=823~nm$. Given that $Si$ is effectively nondispersive over a wide wavelength range\cite{franta2017temperature}, our structure is therefore scalable. Accordingly, the dimensions are normalized to the resonant wavelength of the chiral qBIC, where the period $P_x=0.547\lambda_0$, $P_y=0.972\lambda_0$; the height of bars $H=0.24\lambda_0$; the length of bars $l_1=0.328\lambda_0$, $l_2=0.345\lambda_0$;the width of bars $w=0.122\lambda_0$. The rotation angles of two bars are $\theta_1=\theta_2=10^\circ$.

During the two stages of symmetry breaking, the length and orientation perturbations $\alpha_d$ and $\alpha_o$ are defined as:

\[
\alpha_d=c_1\dfrac{l_2-l_1}{l_2},
\]

\[
\alpha_o=c_2\sin\theta.
\]

The hybrid asymmetry factor is then defined as

\[
\alpha_h=
\begin{cases}
\alpha_d, & 0\le \alpha_h\le 0.5,\\[8pt]
0.5+\alpha_o, & 0.5<\alpha_h\le 1.
\end{cases}
\]

where \(c_1\) and \(c_2\) are chosen such that the normalized length and orientation perturbation span the range \([0,0.5]\).

The polarization state in the far field is evaluated at a port located on a plane normal to the propagation direction, where the far-field complex amplitudes are calculated using\cite{zhen2014topological}: 
\[
C_x=\iint E_x(x,y)e^{-ik_xx-ik_yy}dxdy, 
\]

\[
C_y=\iint E_y(x,y)e^{-ik_xx-ik_yy}dxdy. 
\]
And then ellipticity and the azimuth angle of far field polarization state are calculated based on the two orthogonal complex amplitudes using:

\[
S_0=|C_x|^2+|C_y|^2, 
\]

\[
S_3=-2Im(C_x^*C_y).
\]

$a$,$b$,$c$,$d$ are the integrated intensities of the parity-decomposed field components and are calculated by: 
\[
a=\iint \lvert E_{x,even} \lvert^2 dS , 
\]
\[
b=\iint \lvert E_{x,odd} \lvert^2 dS , 
\]

\[
c=\iint \lvert E_{y,even} \lvert^2 dS,
\]

\[
d=\iint \lvert E_{y,odd} \lvert^2 dS .
\]

\newpage
\subsection*{Note S2: Another evolving route of rectangular bar dimer: rotate first and then length variation}

\begin{figure}[H]
    \centering
    \includegraphics[width=1\linewidth]{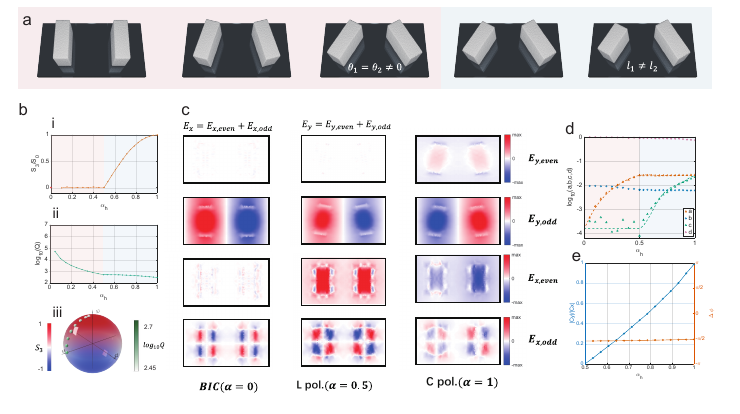}
    \caption{\textbf{Another geometry evolution route from BIC to chiral qBIC for rectangular bar dimer.} (a).Schematic illustration of the geometric evolution of a rectangular dimer. (b). ( \romannumeral 1 ) $S_3/S_0$ and ( \romannumeral 2 ) Q factor evolving with normalized asymmetry factor $\alpha_h$. ( \romannumeral 3 ) The polarization state evolution trajectory on Poincare sphere. Darker green small spheres indicate higher Q factor. (c). Even and odd components of the electric field profiles in the XOY plane for the BIC, linearly polarized (LP) and circularly polarized (CP) states. 
    (d). The intensity of the decomposed electric field evolving with $\alpha_h$.
    (e) The amplitude ratio and phase difference of two orthogonal channels $C_x$ and $C_y$ at far field. } 
    \label{fig:recbar}
\end{figure}

 Fig. \ref{fig:recbar} exhibits the second evolution route of the rectangular bar dimer. Compared to the evolution route in Fig. 2, both routes start from the same BIC eigenmode and end at the same chiral qBIC state. However, the mirror symmetry about the X-axis is first broken by rotating both bars by the same angle $\theta$. Then, the mirror symmetry about the Y-axis is further broken by reducing the length of one of the bars, as shown in the schematic in Fig. \ref{fig:recbar}(a). The second evolution route shows a similar effect on the Q factor and $S_3/S_0$. The Q factor shows an inverse-square relationship with the asymmetry factor $\alpha_h$ in the first stage, and keeps reducing but deviates from the $Q \propto \alpha_{h}^{-2}$ relationship in the second stage.

 In contrast, due to the complementary sequence of symmetry breaking, the far field polarization during evolution is significantly different. As mirror symmetry with respect to X-axis is first broken, the even component of $E_x$ is first generated while $E_y$ is still prohibited($\alpha_h=0.5$ in Fig. \ref{fig:recbar}(c)). As a result, the far field radiation when $S_3/S_0=0(\alpha_h\leq0.5)$ is linearly polarized in the X direction instead of Y direction, which is also quantitatively illustrated in Fig.\ref{fig:recbar}(d), where the evolution of the even symmetry component intensities $a$ and $c$ is opposite to that shown in Fig. 2(d). Moreover, when $\alpha_h>0.5$, this evolution route shows little impact on the azimuthal angle of the polarization state, as shown on the Poincaré sphere in Fig. \ref{fig:recbar}(b)(\romannumeral 3).

\newpage
\subsection*{Note S3: The schematic illustration of geometric evolution of different dimers.}

\begin{figure}[H]
    \centering
    \includegraphics[width=1\linewidth]{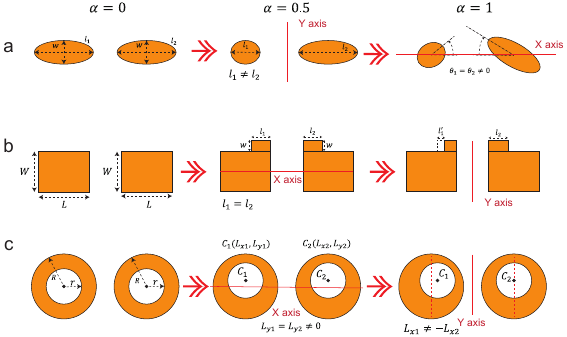}
    \caption{\textbf{Geometric configuration and its evolution with asymmetry perturbation of three kinds of dimers.} (a)elliptical pillars(EP) (b)block bar(BB). (c)cylinder with hole(CH).}
    \label{ddmiers}
\end{figure}

Fig.\ref{ddmiers} exhibits the general symmetry breaking framework for three types of different dimers. When normalized $\alpha_h \leq 0.5$, only a single asymmetry perturbation is introduced, whereas for $\alpha_h > 0.5$, a second asymmetry perturbation breaking mirror symmetry orthogonal to the first one is applied. 

For EP, similar to rectangular bar case, the geometry mirror symmetry about Y axis is first broken by reducing the length of one of the elliptical pillars($\alpha_d=c_1{(l_2-l_1)}/{l_2}$), followed by breaking the symmetry about X axis by rotating the long axis of both elliptical pillars by the same angle($\alpha_o=c_2\sin\theta$). where $c_1$ and $c_2$ normalize the two perturbations terms to the ranges \([0,0.5]\). The same convention is used hereafter. The hybrid asymmetry factor $\alpha_h$ is defined as:

\[
\alpha_h=
\begin{cases}
\alpha_d, & 0\le \alpha_h\le 0.5,\\[8pt]
0.5+\alpha_o, & 0.5<\alpha_h\le 1.
\end{cases}
\]

The resonant wavelength of the chiral qBIC mode($\alpha=1$) of EP is $\lambda_{ep}=931~nm$. The period of the dimer lattice $P_x=0.859\lambda_{ep}$, $P_y=0.466\lambda_{ep}$; the long axis of two bars $l_1=0.097\lambda_{ep}$, $l_2=0.15\lambda_{ep}$; the short axis of both bars $w=0.091\lambda_{ep}$; the height of the elliptical pillars $H=0.43\lambda_{ep}$. The major axes of both bars are rotated by the same angle $\theta_1=\theta_2=6^\circ$.

For BB, the evolution starts with two rectangular blocks, and the mirror symmetry about X axis is broken by adding an extra small block bar on the top of each large block($\alpha_1=c_1l_1/L$). Further, the mirror symmetry about Y axis is broken by reducing the length of one of the small block bars($\alpha_2=c_2l{_1}^{'}/l_1$), as shown in Fig. \ref{ddmiers}(b). The hybrid asymmetry factor is defined as: 

\[
\alpha_h=
\begin{cases}
\alpha_1, & 0\le \alpha_h\le 0.5,\\[8pt]
0.5+\alpha_2, & 0.5<\alpha_h\le 1.
\end{cases}
\]

The resonant wavelength of the chiral qBIC mode is $\lambda_{bb}=1045~nm$. The period of the dimer lattice $P_x=0.67\lambda_{bb}$, $P_y=0.415\lambda_{bb}$; the length and width of the large block $L=0.23\lambda_{bb}$, $W=0.211\lambda_{bb}$; the length of two small bars $l_1=0.048\lambda_{bb}$, $l_2=0.105\lambda_{bb}$; the width of both small bars $w=0.077\lambda_{bb}$; the height of the dimers $H=0.191\lambda_{bb}$.

The symmetry breaking of CH is controlled by the position of its hole. First, the centers of both holes are displaced from the cylinder center along the Y direction($\alpha_y=c_1*L_y/r$), thereby breaking the mirror symmetry about X-axis. Then, the center of one of the holes is shifted along X direction, and thus the mirror symmetry about Y axis is broken($\alpha_x=c_2*L_x/r$).The hybrid asymmetry factor is then defined as: 

\[
\alpha_h=
\begin{cases}
\alpha_y, & 0\le \alpha_h\le 0.5,\\[8pt]
0.5+\alpha_x, & 0.5<\alpha_h\le 1.
\end{cases}
\]

However, although the mirror symmetry is broken in the first stage, the BIC is not broken and we will further discuss it in Fig. S4. The corresponding resonant chiral qBIC wavelength is $\lambda_{ch}=953~nm$. The period of the dimer lattice $P_x=0.839\lambda_{ch}$, $P_y=0.455\lambda_{ch}$; the radius of the cylinder pillar $R=0.157\lambda_{ch}$; the radius of the hole $r=0.079\lambda_{ch}$; offsets of the two hole centers from their respective cylinder centers along the X direction, $L_{x1}=0.034\lambda_{ch}$, $L_{x2}=0~nm$; offset of the two hole centers from their respective cylinder centers along the Y direction, $L_{y1}=L_{y2}=0.044\lambda_{ch}$; the height of the dimers $H=0.421\lambda_{ch}$.

\newpage
\subsection*{Note S4: Parity electric field distribution of cylinder-hole dimer.}

\begin{figure}[H]
    \centering
    \includegraphics[width=1\linewidth]{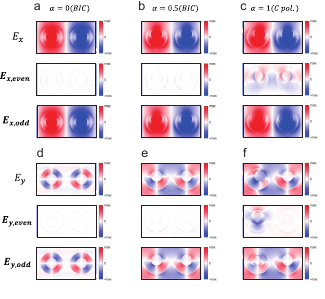}
    \caption{\textbf{The profiles of $E_x$ and $E_y$ in XOY plane and their parity components}}
    \label{chelectric}
\end{figure}

 When no asymmetry perturbation is introduced, the hole center is located at the cylinder center. $E_x$ and $E_y$ possess odd parity symmetry with respect to Y- and X-axes respectively, as denoted by the red dotted lines in Fig.\ref{chelectric}(a) and (d). Notably, as the distance between two cylinders within one lattice is equal to that to the adjacent cylinder, each individual cylinder-hole also possesses odd parity symmetry along the Y direction, as denoted by green dotted line in Fig.\ref{chelectric}(d). As a result, when the two holes shift along the Y direction, the odd parity symmetry of $E_x$ with respect to Y-axis is preserved(Fig.\ref{chelectric}(b)). In contrast, despite the odd parity symmetry breaking of $E_y$ with respect to X-axis, the symmetry for each individual cylinder-hole remains(Fig.\ref{chelectric}(e)), and the far-field radiation is still prohibited by the localized symmetry. Therefore, when $0<\alpha_h \leq 0.5$, as shown in Fig.3(c), the system remains at BIC state with extremely large Q factor and ill-defined polarization. When one of the holes shifts along the X direction subsequently, odd parity symmetries of $E_x$ and $E_y$ are broken simultaneously by this perturbation, with even parity components appearing(Fig.\ref{chelectric}(c)(f)) and finally reaches the circularly polarized state.  

 \newpage
\subsection*{Note S5: Far field polarization components Cx and Cy of different dimers.}

\begin{figure}[H]
    \centering
    \includegraphics[width=1\linewidth]{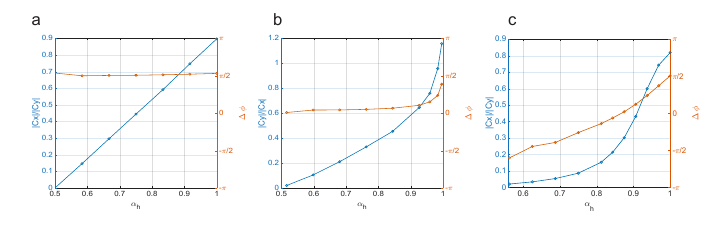}
    \caption{\textbf{The amplitude ratio and phase difference of $C_x$ and $C_y$.}(a)elliptical pillar; (b)block bar; (c) cylinders with off-centered holes.}
    \label{pcmmm}
\end{figure}

As shown in Fig. 3(c), the evolution of the far-field polarization can be understood through the amplitudes and relative phase of the two orthogonal radiative channels. 
During the first perturbation stage $\alpha<0.5$, only one channel is efficiently released, resulting in an amplitude ratio close to zero for all three structures. Once the second perturbation is introduced $\alpha_h>0.5$, the orthogonal channel becomes accessible and the amplitude ratio $\lvert C_x/C_y \rvert$ (or $\lvert C_y/C_x \rvert$ for CH) increases continuously toward unity. This behavior reflects the progressive balancing of the two radiative channels required for circularly polarized emission.

The phase evolution exhibits structure-dependent dynamics. 
For EP, the phase difference $\Delta \phi$ remains close to $\pi/2$ throughout the second stage, indicating that chirality is primarily governed by amplitude balancing. 
In contrast, BB and CH exhibit a gradual evolution of $\Delta \phi$ toward $\pi/2$, suggesting that both amplitude and phase evolution contribute to the formation of the circularly polarized state. 
Despite these differences, all three structures converge toward the same condition for circular polarization, namely comparable amplitudes of the orthogonal radiative channels together with a phase difference approaching $\pi/2$.

\newpage
\subsection*{Note S6: Schematic of elliptical pillar dimers with phase change material $Sb_2S_3$}

\begin{figure}[H]
    \centering
    \includegraphics[width=0.9\linewidth]{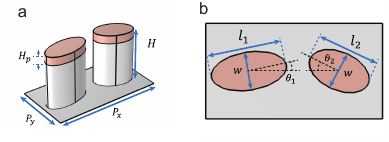}
    \caption{\textbf{The schematic of elliptical pillar dimers integrating with phase change material $Sb_2S_3$.}(a)oblique view (b)top view.}
    \label{pcmmm}
\end{figure}

The geometry regularity can be further broken by allowing the two elliptical pillars to rotate at different angles. The geometry configuration of the elliptical dimer with chiral qBIC mode in Fig. 4(a) is labeled in Fig. \ref{pcmmm}. The resonant wavelength of the chiral qBIC mode($\alpha=1$) is $\lambda_0=978~nm$; the periodic in X and Y direction $P_x=0.82\lambda_0$, $P_y=0.444\lambda_0$; the height of two pillars $H=0.409\lambda_0$; the long axes of two ellipse pillars $l_1=0.344\lambda_0$, $l_2=0.286\lambda_0$; the short axes of two ellipse pillars $w=0.174\lambda_0$. 

The first symmetry breaking perturbation is defined as $\alpha_o'=c\sin(\theta)$. After the first perturbation is introduced, $\theta=\theta_1=\theta_2=24^\circ$. In second symmetry breaking stage,  $\alpha_d=c(l_2-l_1)/l_2$. The third perturbation $\theta^{'}_{1}=-15^\circ$ further decreases $\theta_1$ to $9^\circ$, introducing an extra asymmetry factor $\alpha_o=c\sin(\theta^{'}_{1})$. 

\[
\alpha_h=
\begin{cases}
\alpha_o', & 0\le \alpha_h\le 0.47,\\[8pt]
0.47+\alpha_d, & 0.47<\alpha_h\le 0.7,\\[8pt]
0.7+\alpha_o, & 0.7<\alpha_h\le 1.
\end{cases}
\]

And the coefficient $c$ normalizes $\alpha_h$ to the range of 0 to 1. When integrated with the phase change material, $Sb_2S_3$, whose refractive index is 3.5 in the crystalline phase and 3 in the amorphous phase\cite{dong2019wide}, all geometry parameters remain the same, apart from an additional $Sb_2S_3$ layer on top of elliptical pillars(as donated by brown layer in Fig. \ref{pcmmm}). The holistic height fixed and the thickness of $Sb_2S_3$ layer is $H_p=0.061\lambda_0$. 

\newpage
\subsection*{Note S7: Amplitude and phase evolution of $C_x$ and $C_y$ with asymmetry parameters }

\begin{figure}[H]
    \centering
    \includegraphics[width=0.9\linewidth]{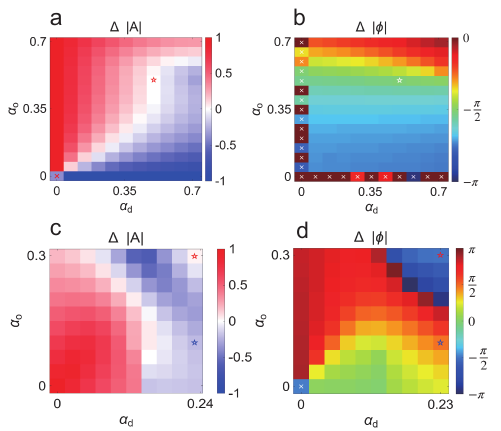}
    \caption{\textbf{Evolution of amplitude and phase difference of $C_x$ and $C_y$ with asymmetry factor. }(a)$\Delta|A|$ and (b) $\Delta\phi$ of rectangular bar dimer. (c)$\Delta|A|$ and (d) $\Delta\phi$ of elliptical pillars with all design degree of freedom released.}
    \label{colormap}
\end{figure}

The far-field radiation polarization is determined by two orthogonal radiative channels, denoted as $C_x$ and $C_y$. When the normalized amplitude difference $\Delta |A|=\dfrac{(|C_x|-|C_y|)}{(|C_x|+|C_y|)}=0$ and the phase difference $\Delta \phi$ reaches $\pm \pi/2$, the radiation becomes perfectly circularly polarized.

For rectangular bar dimer shown in Fig. 2, when $\alpha_o=\alpha_d=0$, the system resides in a BIC state, where both $\Delta |A|$ and $\Delta |\phi|$ are ill-defined. When only one of $\alpha_d$ and $\alpha_o$ is 0, $\Delta |A|$ reaches $\pm 1$ while $\Delta |\phi|$ remains ill-defined, as one of the radiative channels is prohibited. With $\alpha_d$ or $\alpha_o$ gradually increasing from 0, $\Delta |A|$ decreases as the orthogonal channel is gradually released. The two channels reaches balance($\Delta |A|=0$) nearly along the diagonal of Fig. \ref{colormap}(a). However, the radiation doesn't reach circularly polarization until $\Delta |\phi|$ is $-\pi/2$(white star in Fig. \ref{colormap}(b)). It is notable $\alpha_d$(the length of rectangular bar) has little affect on  $\Delta |\phi|$, while $\alpha_o$(the rotated angle) is able to monotonically tune the phase difference.   

For ellipse pillar dimer shown in Fig. 4, as shown in Fig. \ref{colormap}(c)(d), the asymmetry factors exhibit substantial modulation of both $\Delta |\phi|$ and $\Delta |A|$. When  $\alpha_d=0.23, \alpha_o=0.3$, $\Delta |A|$ decreases to 0 with a $\Delta |\phi|$ of $\dfrac{\pi}{2}$, leading to perfect circularly polarization(denoted by red star). It is also observed that variations in $\alpha_o$ have limited influence on $\Delta |A|$ when $\alpha_d=0.23$. However, $\Delta |\phi|$ is significantly modulated by $\alpha_o$ and decreases from $\dfrac{\pi}{2}$ to  $-\dfrac{\pi}{2}$(denoted by blue star), resulting in a handedness reversal of circularly polarization. This also reveals that the significant reversal of $S_3/S_0$ from -0.91 to 1 in Fig. 4(b)(\romannumeral 1) is attributed to the reversal of phase difference.  

\newpage
\subsection*{Note S8: The trade-off between Q-factor and chirality}

\begin{figure}[H]
    \centering
    \includegraphics[width=1\linewidth]{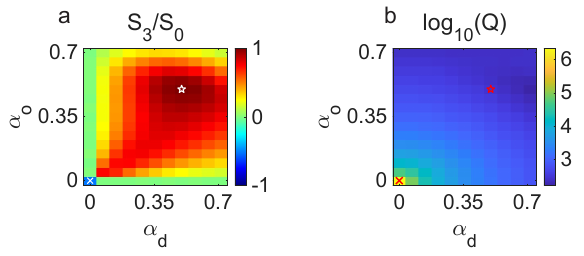}
    \caption{\textbf{Evolution of (a) $S_3/S_0$ and (b) Q factor with asymmetry factor of rectangular bar dimer.} }
    \label{colormap2}
\end{figure}

As shown in Fig. S8, in the absence of symmetry breaking($\alpha_o=\alpha_d=0$), the system remains in the BIC state with $S_3/S_0$ ill-defined and infinite Q factor. Introducing a single asymmetry perturbation would eliminate the BIC state($\alpha_{o/d}=0$) while S3 is still constrained to 0. Once both orthogonal symmetry-breaking perturbations are introduced($\alpha_d \neq 0, \alpha_o \neq 0$), chirality emerges immediately as indicated by a non-zero $S_3/S_0$ and finally peaks at ($\alpha_d=\alpha_o=0.5$).

Notably, although chirality can be easily obtained with most combinations of $\alpha_o$ and $\alpha_d$, the ellipticity does not always reach its maximum($S_3/S_0$=1). As depicted in Fig.\ref{colormap}(a)(b), maximum chirality is acheived at the intersection of zero amplitude difference and a phase difference of $\pm \pi/2$.

In contrast, the Q factor decreases monotonically with increasing asymmetry factor owing to enhanced radiative leakage, leading to an inevitable trade-off between Q factor and $S_3/S_0$. Nevertheless, since the evolution originates from a BIC eigenmode, the Q factor remains as high as 300 when chirality reaches its maximum(red star in Fig.\ref{colormap2}(b)).

\newpage
\bibliography{Suppl}